\documentclass[11pt]{article}
\usepackage[margin=1in]{geometry}
\usepackage{amsmath, amssymb, amsthm}
\usepackage{graphicx}
\usepackage{siunitx}
\DeclareSIUnit{\atm}{atm}
\DeclareSIUnit{\mmHg}{mmHg}
\DeclareSIUnit{\ppm}{ppm}
\DeclareSIUnit{\bpm}{bpm}
\usepackage{cite}
\usepackage{booktabs}
\usepackage{tabularx}
\usepackage{algorithm}
\usepackage[hidelinks]{hyperref}
\usepackage{algpseudocode}
\usepackage{tikz}
\usetikzlibrary{arrows.meta, positioning, shapes.geometric, fit, calc}
\usepackage{pgfplots}
\pgfplotsset{compat=1.18}
\usepackage{float}
\usepackage{enumitem}
\usepackage{xcolor}
\usepackage{cleveref}
\newtheorem{remark}{Remark}
\title{\textbf{Physicochemical-Neural Fusion for Semi-Closed-Circuit Respiratory Autonomy in Extreme Environments}}
\author{
\vspace{1em}
\begin{minipage}[t]{0.5\textwidth}
    \centering
    \begin{tabular}[t]{c}
        \small{\textbf{Phillip Kingston}\thanks{Corresponding author: phillip.kingston@galacticbioware.com}}\\ \vspace{-0.4em}
        \scriptsize{Member of Technical Staff} \\
        \scriptsize{Galactic Bioware} \\
    \end{tabular}
\end{minipage}%
\begin{minipage}[t]{0.5\textwidth}
    \centering
    \begin{tabular}[t]{c}
        \small{\textbf{Nicholas Johnston}} \\ \vspace{-0.4em}
        \scriptsize{Member of Technical Staff} \\
        \scriptsize{Galactic Bioware} \\
    \end{tabular}
\end{minipage}%
\vspace{1em}
}
\date{\normalsize{1 March 2026}}
\begin{document}
\maketitle
\begin{center}
    \textbf{CC BY-NC-SA 4.0} \\
    \vspace{0.3em}
    \footnotesize{This work is licensed under a Creative Commons Attribution-Noncommercial-ShareAlike 4.0 International License (CC BY-NC-SA 4.0)}
\end{center}
\begin{abstract}
This paper introduces \emph{Galactic Bioware's Life Support System}, a semi-closed-circuit breathing apparatus designed for integration into a positive-pressure firefighting suit and governed by an AI control system. The breathing loop incorporates a soda lime CO\textsubscript{2} scrubber, a silica gel dehumidifier, and pure O\textsubscript{2} replenishment with finite consumables. One-way exhaust valves maintain positive pressure while creating a semi-closed system in which outward venting gradually depletes the gas inventory. \textbf{Part~I} develops the physicochemical foundations from first principles, including state-consistent thermochemistry, stoichiometric capacity limits, adsorption isotherms, and oxygen-management constraints arising from both fire safety and toxicity. \textbf{Part~II} introduces an AI control architecture that fuses three sensor tiers---external environmental sensing, internal suit atmosphere sensing (with triple-redundant O\textsubscript{2} cells and median voting), and firefighter biometrics. The controller combines receding-horizon model-predictive control (MPC) with a learned metabolic model and a reinforcement learning (RL) policy advisor, with all candidate actuator commands passing through a final control-barrier-function safety filter before reaching the hardware. This architecture is intended to optimize performance under unknown mission duration and exertion profiles. In this paper:
\begin{enumerate}
\item We introduce a rigorous first-principles physicochemical model of the semi-closed breathing loop, including state-consistent thermochemistry of soda lime scrubbing, GAB-isotherm humidity management, and the oxygen-enrichment dynamic driven by exhaust-valve vent compensation with pure O\textsubscript{2}, subject to fire-safety constraints.
\item We introduce an 18-state, 3-control nonlinear state-space formulation using only sensors viable in structural firefighting, with triple-redundant O\textsubscript{2} sensing and median voting.
\item We introduce an MPC framework with a dynamic resource scarcity multiplier, an RL policy advisor for warm-starting, and a final control-barrier-function safety filter through which all actuator commands must pass, demonstrating 18--34\% endurance improvement in simulation over PID baselines while maintaining tighter physiological and fire-safety margins.
\end{enumerate}
\end{abstract}
\newpage
\tableofcontents
\newpage
\part{Chemical and Physical Foundations}
\section{Introduction}
\label{sec:introduction}
Firefighters operate in environments filled with smoke, toxic gases, and extreme temperatures, often exceeding \SI{500}{\celsius} in structural fires. Conventional open-circuit self-contained breathing apparatus (SCBA) exhaust each breath to the environment, wasting roughly two-thirds of the delivered oxygen and limiting operating time to approximately 30 minutes under heavy exertion \cite{lightweight, tactical, physiological}. A closed-circuit breathing apparatus (CCBA) recycles exhaled gas, scrubs CO\textsubscript{2}, removes excess moisture, and replenishes consumed O\textsubscript{2} from a finite supply. Although this approach is well established in military diving and spacecraft life support \cite{nasa, mixedgas}, fatal CO\textsubscript{2} retention incidents in rebreather diving demonstrate that it is not without risk and depends critically on reliable scrubbing \cite{extremepressure}. If implemented safely, it can potentially triple effective operating time while reducing the firefighter’s overall carried weight. This paper makes two contributions. \textbf{Part~I} develops the chemical and physical foundations of the Galactic Bioware Life Support System from first principles, including complete thermochemical analyses, adsorption theory, and airflow dynamics. \textbf{Part~II} introduces an AI-based control system that uses sensor fusion and online optimization to manage finite consumables against unknown mission duration and dynamically evolving fireground conditions.
The system is \emph{semi-closed}: the positive-pressure suit incorporates one-way exhaust valves (consistent with NFPA 1991 Level~A encapsulating suit practice) that vent gas outward when internal pressure exceeds a cracking pressure, preventing toxic infiltration while allowing controlled pressure relief. This intermittent venting creates a slow net loss of gas from the suit, which is compensated by O\textsubscript{2} injection---the primary driver of oxygen enrichment and the central control challenge addressed by the AI system.
The system emphasizes three design constraints:
\begin{enumerate}[label=(\roman*)]
    \item \textbf{Semi-closed positive pressure:} The suit vents outward through exhaust valves but never admits external air. Venting depletes the gas inventory, requiring make-up O\textsubscript{2} injection that drives gradual oxygen enrichment.
    \item \textbf{Variable metabolic demand:} Oxygen consumption is a nonlinear function of firefighter exertion, thermal stress, and psychological state.
    \item \textbf{Finite consumables:} System endurance is bounded by limited soda lime (\SI{1}{\kilogram}), silica gel (\SI{1}{\kilogram}), and oxygen supply (\SI{3}{\kilogram}).
\end{enumerate}
\section{System Overview}
\label{sec:system_overview}
\subsection{Components and Layout}
The Galactic Bioware Life Support System comprises the following subsystems:
\begin{enumerate}[label=(\roman*)]
    \item \textbf{Positive-Pressure Suit with Exhaust Valves:} The suit maintains an internal gauge pressure $\Delta P_{\mathrm{suit}} = P_s - P_a > 0$ (typically \SIrange{2}{5}{\milli\bar}) relative to the ambient pressure $P_a$. One-way exhaust valves, consistent with NFPA 1991 Level~A encapsulating suit design, open at a cracking pressure $P_{\mathrm{crack}} \approx P_a + \SI{5}{\milli\bar}$ to vent gas outward, preventing overpressure while ensuring that any leak pathway results in outward gas flow. This makes the suit \emph{semi-closed}: gas is never admitted from the environment, but is intermittently vented outward.
    \item \textbf{Air Circulation System:} Two variable-speed brushless DC fans---an \emph{outtake} fan drawing exhaled air from the suit interior and an \emph{intake} fan returning treated air---drive a continuous flow through the treatment train.
    \item \textbf{CO\textsubscript{2} Scrubber:} A packed-bed canister of granular soda lime (Ca(OH)\textsubscript{2}/NaOH formulation, mean granule diameter \SIrange{2}{5}{\milli\meter}, with pH-indicating dye) removes exhaled CO\textsubscript{2} through irreversible acid--base neutralization. Soda lime is the standard sorbent in closed-circuit breathing systems (rebreathers, anesthesia circuits) because its calcium hydroxide matrix binds the caustic alkali, preventing the formation of free NaOH solution that could cause airway burns. This stage is positioned \emph{first} in the treatment train to receive moist exhaled gas directly, maintaining the aqueous surface film required for efficient scrubbing (see \cref{sec:airflow_pathway}).
    \item \textbf{Dehumidification Unit:} A packed-bed canister of indicating silica gel (Type~A, mean bead diameter \SIrange{2}{5}{\milli\meter}) removes water vapor from the circulating gas stream via physical adsorption. Positioned \emph{downstream} of the scrubber, it captures both exhaled moisture and reaction-generated moisture in a single pass.
    \item \textbf{Oxygen Replenishment System:} A proportional solenoid valve meters gaseous O\textsubscript{2} from a high-pressure composite tank containing \SI{3.0}{\kilogram} O\textsubscript{2} ($\approx$\SI{93.75}{\mole}, $\approx$\SI{2100}{\liter} at STP), stored at \SI{200}{\bar} in a $\sim$\SI{11.7}{\liter} cylinder. The tank mass accounts for non-ideal gas behavior at high pressure. Using the real-gas equation:

\begin{equation}
n = \frac{P V}{Z R T}
\end{equation}

with compressibility factor $Z \approx 0.95$ for oxygen at \SI{200}{\bar} 
and \SI{300}{\kelvin} (per NIST thermophysical properties data for 
O\textsubscript{2}), the total gas mass at fill is approximately 
\SI{3.16}{\kilogram}. Because $Z < 1$ at this pressure, the cylinder 
holds \emph{more} gas than an ideal-gas estimate would predict. However, 
the regulator requires a minimum inlet pressure of \SIrange{10}{25}{\bar} 
to maintain stable delivery, below which the residual gas 
(\SIrange{0.15}{0.38}{\kilogram}) is unrecoverable. The design therefore 
adopts a nominal \emph{usable} capacity of \SI{3.0}{\kilogram}, 
corresponding to a minimum delivery pressure of approximately 
\SI{11}{\bar}.
    \item \textbf{Counter-Lung (Breathing Bag):} A flexible bellows or collapsible bag connected to the breathing loop that accommodates tidal breathing oscillations and transient mismatches between O\textsubscript{2} injection and metabolic consumption at near-constant pressure.
    \item \textbf{Sensor Suite and AI Controller:} Three categories of sensors---external environmental, internal suit environment (including triple-redundant O\textsubscript{2} cells), and firefighter biometric---feed the AI control system described in detail in Part~II.

\end{enumerate}
\subsection{Airflow Pathway}
\label{sec:airflow_pathway}
The ordering of treatment stages in the closed loop is dictated by a critical physical-chemistry constraint: the soda lime scrubbing reaction \emph{requires} moisture to proceed, while simultaneously \emph{producing} moisture as a byproduct. This coupling determines the optimum topology.

\begin{figure}[H]
    \centering
    \includegraphics[width=0.95\textwidth]{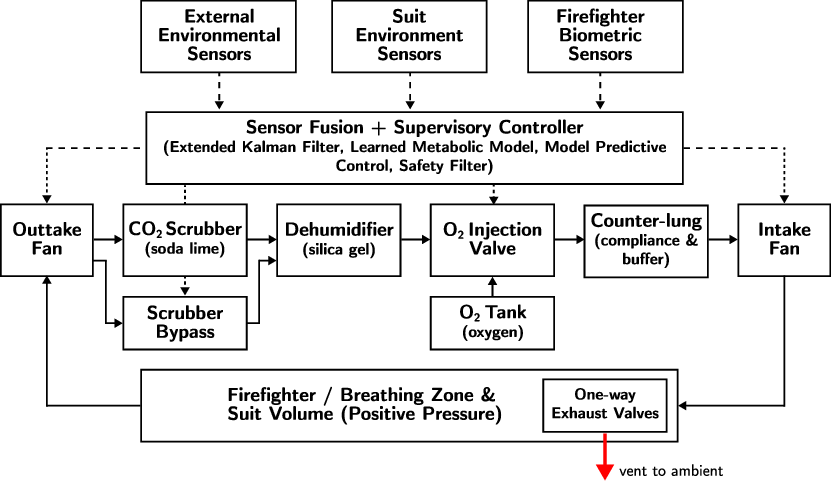}
    \caption{Control-oriented architecture of the semi-closed positive-pressure breathing loop. Solid lines illustrate gas flow and dashed lines sensor inputs / control links. Moist exhaled gas is drawn through the soda-lime scrubber first, then the downstream silica-gel dehumidifier, before O$_2$ replenishment and return to the breathing zone. Outward venting through one-way exhaust valves makes the suit semi-closed; the controller uses external, in-suit, and biometric sensing to regulate O$_2$ injection, fan speed, and scrubber bypass.}
    \label{fig:system_schematic_control}
\end{figure}

\subsubsection{Gas-Treatment Stage Ordering}
One possible configuration places the dehumidifier upstream of the scrubber; however, this topology is incompatible with the underlying physicochemical requirements for two reasons:

\begin{enumerate}[label=(\roman*)]
    \item \textbf{Scrubber moisture requirement:} As detailed in \cref{sec:co2}, the CO\textsubscript{2} absorption mechanism proceeds through an aqueous surface film on the soda lime granules (\cref{eq:henry}--\cref{eq:neut2}). If the inlet gas is aggressively dried, this film desiccates, causing a hard crust of calcium carbonate to form on the granule surface. The crust blocks gas--liquid contact, and the effectiveness factor $\eta(t)$ drops precipitously---potentially rendering the scrubber non-functional while significant hydroxide remains unreacted in the granule interior.

    \item \textbf{Reaction-generated moisture:} Each mole of CO\textsubscript{2} scrubbed produces one mole of H\textsubscript{2}O (\cref{eq:scrub_primary}). At heavy exertion ($\dot{n}_{\mathrm{CO_2}} \approx \SI{0.068}{\mole\per\minute}$), this corresponds to a water production rate of:
    \begin{equation}
    \label{eq:scrub_water}
    \dot{m}_{\mathrm{H_2O,rxn}} = \dot{n}_{\mathrm{CO_2}} \times M_{\mathrm{H_2O}} = 0.068 \times 18.015 \approx \SI{1.22}{\gram\per\minute}
    \end{equation}
    If the desiccant is upstream, this reaction-generated moisture passes \emph{unremoved} into the breathing gas delivered to the firefighter. Combined with the moisture that the upstream desiccant failed to intercept (because it was already saturated or because the scrubber added new moisture downstream), the suit humidity constraint (RH~$\le 60\%$) would be violated within minutes under heavy exertion.
\end{enumerate}
\subsubsection{Gas Pathway}
The ideal closed-loop gas pathway therefore proceeds as follows:
\begin{enumerate}
    \item \textbf{Outtake fan:} Exhaled gas (enriched in CO\textsubscript{2} and H\textsubscript{2}O, depleted in O\textsubscript{2}) is drawn from the suit helmet and torso plenum.
    \item \textbf{Soda lime scrubber:} The \emph{moist} exhaled gas enters the scrubber directly. The high humidity of exhaled air (\SIrange{90}{100}{\percent} RH at \SI{34}{\celsius}) maintains the aqueous surface film on the soda lime granules, ensuring efficient gas--liquid contact for CO\textsubscript{2} absorption. The scrubber simultaneously removes CO\textsubscript{2} and generates additional H\textsubscript{2}O and heat.
    \item \textbf{Silica gel dehumidifier:} The cooled gas, now carrying \emph{both} the firefighter's exhaled moisture and the scrubber's reaction-generated moisture, passes through the silica gel canister. Because this stage sees the \emph{total} system moisture load, the desiccant is utilized efficiently---it captures all sources of water in a single pass rather than missing the largest contributor. However, water adsorption onto silica gel is itself exothermic: the enthalpy of adsorption comprises the latent heat of condensation ($\sim$\SI{2440}{\kilo\joule\per\kilogram}) plus the excess surface energy of sorption ($\sim$\SIrange{100}{200}{\kilo\joule\per\kilogram}), totaling approximately \SI{2550}{\kilo\joule\per\kilogram} of water adsorbed. At the peak moisture load of \SI{4.2}{\gram\per\minute} (\cref{sec:moisture_budget}), this produces:
    \begin{equation}
    \label{eq:heat_ads}
    \dot{Q}_{\mathrm{ads}}(t) \approx 2550 \times \frac{4.2 \times 10^{-3}}{60} \approx \SI{179}{\watt}
    \end{equation}
    This heat is released post-scrubber directly into the gas stream approaching the firefighter's breathing zone. Without a secondary cooling stage, inspired gas temperature could exceed \SI{45}{\celsius}---creating a ``hot hair dryer'' effect that accelerates core temperature rise and thermal injury to the upper airway.
    \item \textbf{Oxygen injection:} Fresh O\textsubscript{2} is metered into the cooled, dehumidified gas stream via the proportional valve.
    \item \textbf{Intake fan:} The refreshed gas (low CO\textsubscript{2}, controlled humidity, replenished O\textsubscript{2}, cooled below \SI{35}{\celsius}) is circulated back into the suit through distribution channels in the helmet, torso, and limbs.
\end{enumerate}
The Galactic Bioware Life Support System is \emph{semi-closed}: it follows the ideal closed-loop gas pathway, modified only by one-way exhaust valves that vent outward intermittently when internal pressure exceeds the cracking pressure, while preventing ambient gas ingress. The net effect is a slow loss of gas-phase moles from the suit, compensated by O\textsubscript{2} injection. This venting is the primary mechanism driving oxygen enrichment in the breathing loop (see \cref{sec:o2_enrichment}).
\subsubsection{Moisture Budget Analysis}
\label{sec:moisture_budget}
To verify that the topology satisfies the humidity constraint, we examine the total moisture load on the downstream desiccant. The two sources of water entering the silica gel canister are:
\begin{enumerate}[label=(\roman*)]
    \item \textbf{Exhaled moisture:} A firefighter under heavy exertion exhales approximately \SIrange{1.5}{3.0}{\gram\per\minute} of water vapor (depending on minute ventilation and body temperature).
    \item \textbf{Scrubber reaction product:} From \cref{eq:scrub_water}, approximately \SI{1.2}{\gram\per\minute} at heavy exertion.
    \item \textbf{Total moisture load:} \SIrange{2.7}{4.2}{\gram\per\minute}, or \SIrange{160}{250}{\gram\per\hour}. 
\end{enumerate}

With \SI{1}{\kilogram} of silica gel at a capacity of $\sim$\SI{350}{\gram} of water, the desiccant provides \SIrange{1.4}{2.2}{\hour} of humidity control at heavy exertion---well-matched to the scrubber and oxygen supply endurance. In the incorrect (upstream) topology, the desiccant would capture only \emph{Exhaled moisture} while \emph{Scrubber reaction product} passes directly to the firefighter, making the humidity constraint unachievable regardless of desiccant capacity.
\section{Carbon Dioxide Management}
\label{sec:co2}
\subsection{Physiological Context}
Normal atmospheric CO\textsubscript{2} concentration is approximately \SI{0.04}{\percent} by volume (\SI{400}{\ppm}). The human body continuously produces CO\textsubscript{2} as the terminal product of aerobic cellular respiration, transported via the blood to the lungs and exhaled at concentrations of \SIrange{3.5}{5.5}{\percent}. In a closed breathing loop, CO\textsubscript{2} accumulates unless actively removed.
The system must maintain inspired CO\textsubscript{2} below 0.5\% (5000\,ppm).
This threshold derives from OSHA's permissible exposure limit
(PEL-TWA, 29 CFR 1910.1000, Table Z-1) \cite{osha_o2}, which is
an 8-hour time-weighted average for occupational ambient air.
A closed-loop breathing apparatus is not an occupational
workspace in the regulatory sense, and no dedicated standard
exists for inspired CO\textsubscript{2} in self-contained life support
systems for firefighting. We adopt the OSHA PEL as a
\emph{conservative operational ceiling} for three reasons:
(i)~firefighter deployments on a single air fill are
substantially shorter than 8 hours, so the TWA averaging
period is never approached;
(ii)~the PEL aligns with the NIOSH recommended exposure limit
(REL) of 5000\,ppm TWA and with submarine atmosphere quality
standards (NAVSEA SS521-AK-HBK-010), providing cross-domain
consistency;
and (iii)~the MPC treats this as a \emph{soft} ceiling with
quadratic penalty onset well below 5000\,ppm (nominal target
$\sim$2000\,ppm), so the controller actively minimises inspired
CO\textsubscript{2} rather than dwelling at the limit.
The hard emergency floor in cascade-failure mode
(\cref{sec:emergency}) is set at 3\%, consistent with
short-duration acute exposure guidance.
\begin{remark}
\SI{1}{\mmHg} is the pressure exerted by a \SI{1}{\milli\meter} column of mercury at \SI{0}{\celsius} under standard gravity, equal to approximately \SI{133.322}{\pascal}.
\end{remark}
\subsection{Performance Degradation from Hypercapnia}
Excessive CO\textsubscript{2} exposure causes \emph{hypercapnia}---an elevation of $P_a\mathrm{CO}_2$ above \SI{45}{\mmHg}. The dissolved CO\textsubscript{2} reacts with water in the blood to form carbonic acid:
\begin{equation}
\label{eq:carbonic}
\mathrm{CO_2(aq)} + \mathrm{H_2O(l)} \rightleftharpoons \mathrm{H_2CO_3(aq)} \rightleftharpoons \mathrm{H^+(aq)} + \mathrm{HCO_3^-(aq)}
\end{equation}
This equilibrium is catalyzed by the enzyme carbonic anhydrase (with a turnover number of $\sim 10^6$~s$^{-1}$) in red blood cells, making the response nearly instantaneous. The resulting increase in $[\mathrm{H^+}]$ lowers blood pH, a condition termed \emph{respiratory acidosis}. The Henderson--Hasselbalch equation quantifies this relationship:
\begin{equation}
\label{eq:henderson}
\mathrm{pH} = \mathrm{p}K_a + \log_{10}\!\left(\frac{[\mathrm{HCO_3^-}]}{[\mathrm{H_2CO_3}]}\right) = 6.1 + \log_{10}\!\left(\frac{[\mathrm{HCO_3^-}]}{0.03 \times P_a\mathrm{CO}_2}\right)
\end{equation}
where $P_a\mathrm{CO}_2$ is measured in mmHg and $0.03$ is the solubility coefficient of CO\textsubscript{2} in plasma (\si{\milli\mole\per\liter} per mmHg) at \SI{37}{\celsius}. Normal arterial pH is \numrange{7.35}{7.45}; respiratory acidosis drives pH below 7.35.
Symptoms progress with severity: at inspired CO\textsubscript{2} of 
\SIrange{2}{3}{\percent}, headache and impaired judgment; at 
\SIrange{5}{7}{\percent}, confusion, tachycardia, and dyspnea; above 
\SI{10}{\percent}, loss of consciousness and death within minutes 
\cite{lambertsen_co2}.
\subsection{Chronic Health Effects}
Repeated subacute exposure produces chronic hypercapnia with renal compensation (elevated serum $\mathrm{HCO_3^-}$), placing sustained strain on the cardiovascular system and increasing hypertension risk. Neurological effects of chronic hypercapnia include impaired memory, concentration deficits, and accelerated cognitive decline. In closed-circuit breathing environments specifically, repeated CO\textsubscript{2} exposure alters respiratory drive and reduces CO\textsubscript{2} sensitivity \cite{co2insensitivity}. Repeated occupational exposure to elevated CO\textsubscript{2} in closed-circuit breathing environments has also been associated with blunted ventilatory chemosensitivity \cite{ventilatoryresponse}, which may mask early warning signs of scrubber failure in experienced users. This observation strengthens the requirement for direct instrumented monitoring of scrubber performance, because physiological perception of rising inspired CO\textsubscript{2} may be attenuated in experienced users.
\subsection{CO\textsubscript{2} Scrubbing with Soda Lime}
\subsubsection{Sorbent Choice and Safety}
The Galactic Bioware Life Support System uses \emph{soda lime}---a granular formulation of calcium hydroxide ($\mathrm{Ca(OH)_2}$, $\sim$75--80\% by mass) with a small fraction of sodium or potassium hydroxide ($\sim$3--5\%) as an activator, plus water ($\sim$15--20\%) and a pH-indicating dye. Soda lime is the standard CO\textsubscript{2} absorbent in closed-circuit breathing systems (military and recreational rebreathers, anesthesia circuits, submarine atmosphere management) because:
\begin{enumerate}[nosep]
    \item The calcium hydroxide matrix physically binds the alkali hydroxide, preventing formation of free caustic solution that could migrate to the breathing zone and cause chemical burns to the airway.
    \item The granular form maintains structural integrity throughout the reaction, with less tendency to form fine particulate than pure NaOH pellets.
    \item The built-in moisture content maintains the aqueous surface film required for efficient gas--liquid contact without relying on external humidity alone.
\end{enumerate}
\begin{remark}[Why not pure NaOH?]
Pure sodium hydroxide pellets are highly hygroscopic and corrosive (pH~13--14 in solution). In a breathing loop, the combination of high humidity, mechanical vibration, and exothermic reaction can produce a mobile caustic liquid---the ``caustic cocktail'' documented in rebreather diving incident reports. This represents an unacceptable airway injury hazard for firefighting. Soda lime mitigates this risk through the calcium hydroxide matrix and controlled formulation.
\end{remark}
\subsubsection{Primary Reaction and Thermochemistry}
The CO\textsubscript{2} absorption in soda lime proceeds through a two-stage mechanism. The NaOH activator reacts first (faster kinetics) and is regenerated by the bulk Ca(OH)\textsubscript{2}:

\medskip
\noindent\textbf{Stage 1: NaOH-catalyzed absorption:}
\begin{align}
\mathrm{CO_2(g) + 2\,NaOH(aq)} &\longrightarrow \mathrm{Na_2CO_3(aq) + H_2O(l)} \label{eq:sl_stage1}
\end{align}

\medskip
\noindent\textbf{Stage 2: Regeneration of NaOH by Ca(OH)\textsubscript{2}:}
\begin{align}
\mathrm{Na_2CO_3(aq) + Ca(OH)_2(s)} &\longrightarrow \mathrm{CaCO_3(s) + 2\,NaOH(aq)} \label{eq:sl_stage2}
\end{align}
The NaOH is recycled; however, because the regeneration step (\cref{eq:sl_stage2}) is slower than the initial absorption (\cref{eq:sl_stage1}), the NaOH can become transiently depleted at high CO\textsubscript{2} loading rates, temporarily reducing the effective scrubbing rate until the Ca(OH)\textsubscript{2} regeneration catches up. The \textbf{net reaction} is:
\begin{equation}
\label{eq:scrub_primary}
\mathrm{CO_2(g) + Ca(OH)_2(s) \longrightarrow CaCO_3(s) + H_2O(l)}
\end{equation}

\medskip
\noindent\textbf{State-consistent enthalpy via formation enthalpies:}
Using standard enthalpies of formation at \SI{298.15}{\kelvin}:
\begin{center}
\begin{tabular}{lS[table-format=-4.1]}
\toprule
\textbf{Species} & {\textbf{$\Delta H_f^\circ$ (\si{\kilo\joule\per\mole})}} \\
\midrule
$\mathrm{CO_2(g)}$ & -393.5 \\
$\mathrm{Ca(OH)_2(s)}$ & -986.1 \\
$\mathrm{CaCO_3(s)}$ (calcite) & -1206.9 \\
$\mathrm{H_2O(l)}$ & -285.8 \\
\bottomrule
\end{tabular}
\end{center}
Applying Hess's law with well-defined initial and final states:
\begin{align}
\Delta H_{\mathrm{rxn}}^\circ &= \bigl[\Delta H_f^\circ(\mathrm{CaCO_3(s)}) + \Delta H_f^\circ(\mathrm{H_2O(l)})\bigr] - \bigl[\Delta H_f^\circ(\mathrm{CO_2(g)}) + \Delta H_f^\circ(\mathrm{Ca(OH)_2(s)})\bigr] \nonumber \\
&= [(-1206.9) + (-285.8)] - [(-393.5) + (-986.1)] \nonumber \\
&= -1492.7 - (-1379.6) \nonumber \\
&= \SI{-113.1}{\kilo\joule\per\mole\text{ CO}_2} \label{eq:enthalpy}
\end{align}
This value is \emph{uniquely determined} by the standard-state endpoints of the net reaction and is path-independent (Hess's law). No value range is needed for the standard enthalpy itself---uncertainty in the operating heat release arises from non-standard conditions (elevated temperature, varying hydration state of the product) and is addressed through engineering safety factors in the thermal management design rather than by adjusting the thermodynamic value.
For comparison, the analogous reaction with pure NaOH yields $\Delta H_{\mathrm{rxn}}^\circ \approx \SI{-171}{\kilo\joule\per\mole}$ if water condenses to liquid, or $\SI{-127}{\kilo\joule\per\mole}$ if water remains vapor. The soda lime reaction is less exothermic per mole of CO\textsubscript{2}, which is a thermal management advantage.
\subsubsection{Reaction Mechanism and Kinetics}
The gas--solid reaction proceeds through a multi-step mechanism:

\medskip
\noindent\textbf{Step 1: Dissolution of CO\textsubscript{2} into the surface water film:} Soda lime granules maintain a thin aqueous film due to their built-in moisture content ($\sim$15--20\% water). In pure water, CO\textsubscript{2} dissolves according to Henry's law:
\begin{equation}
\label{eq:henry}
[\mathrm{CO_2(aq)}] = K_H \cdot p_{\mathrm{CO_2}}
\end{equation}
where $K_H = 3.4 \times 10^{-2}$~mol\,L$^{-1}$\,atm$^{-1}$ at \SI{298}{\kelvin} for pure water (solubility convention; $[\mathrm{CO_2}] = K_H \cdot p$). However, the soda lime surface film is a concentrated alkaline solution (pH~12--14), in which dissolved CO\textsubscript{2} is rapidly consumed by reaction with OH$^-$ (Steps~2--3). The result is \emph{reaction-enhanced absorption}: the effective gas-phase driving force is much larger than the bare Henry's law equilibrium would predict, because the liquid-side CO\textsubscript{2} concentration is held near zero by fast chemical consumption. This regime is characterized by a Hatta number $\mathrm{Ha} \gg 1$, and the absorption rate is governed by the product $K_H \sqrt{k_{\mathrm{OH}} [\mathrm{OH}^-] D_{\mathrm{CO_2}}}$ rather than by $K_H$ alone, where $k_{\mathrm{OH}}$ is the second-order rate constant for the CO\textsubscript{2}$+$OH$^-$ reaction and $D_{\mathrm{CO_2}}$ is the liquid-phase diffusivity. The overall volumetric scrubbing rate (\cref{eq:scrub_rate}) absorbs this enhancement into the lumped coefficient $k_{\mathrm{ov}}$.

\medskip
\noindent\textbf{Step 2: Formation of carbonic acid and dissociation:}
\begin{equation}
\mathrm{CO_2(aq)} + \mathrm{H_2O(l)} \rightleftharpoons \mathrm{H_2CO_3(aq)} \rightleftharpoons \mathrm{H^+(aq)} + \mathrm{HCO_3^-(aq)}
\end{equation}
with $K_{a1} = \num{4.3e-7}$ at \SI{298}{\kelvin}.

\medskip
\noindent\textbf{Step 3: Neutralization by hydroxide ions:}
\begin{align}
\mathrm{H_2CO_3(aq)} + \mathrm{OH^-(aq)} &\longrightarrow \mathrm{HCO_3^-(aq)} + \mathrm{H_2O(l)} \label{eq:neut1} \\
\mathrm{HCO_3^-(aq)} + \mathrm{OH^-(aq)} &\longrightarrow \mathrm{CO_3^{2-}(aq)} + \mathrm{H_2O(l)} \label{eq:neut2}
\end{align}
\noindent\textbf{Step 4: Precipitation of calcium carbonate:}
\begin{equation}
\mathrm{Ca^{2+}(aq) + CO_3^{2-}(aq) \longrightarrow CaCO_3(s)}
\end{equation}
The rate-limiting step under typical operating conditions is the gas-phase mass transfer of CO\textsubscript{2} to the granule surface (Step~1), which depends on the gas-phase velocity, granule surface area, and the driving force $p_{\mathrm{CO_2,bulk}} - p_{\mathrm{CO_2,surface}}$. We model the overall volumetric rate of CO\textsubscript{2} removal as:
\begin{equation}
\label{eq:scrub_rate}
r_{\mathrm{scrub}} = k_{\mathrm{ov}}\,a_s\,V_{\mathrm{bed}}\,\bigl(p_{\mathrm{CO_2}} - p_{\mathrm{CO_2}}^*\bigr)\,\eta(t)
\end{equation}
where $k_{\mathrm{ov}}$ is the overall mass transfer coefficient (\si{\mole\per\second\per\pascal\per\meter\squared}), $a_s$ is the specific surface area of the packed bed (\si{\meter\squared\per\meter\cubed}), $V_{\mathrm{bed}}$ is the bed volume, $p_{\mathrm{CO_2}}^*$ is the equilibrium partial pressure over the product layer (effectively zero for fresh soda lime), and $\eta(t) \in [0,1]$ is an effectiveness factor that decreases as the Ca(OH)\textsubscript{2} is consumed and the product layer of $\mathrm{CaCO_3}$ builds up on granule surfaces, creating a diffusion barrier.
\subsubsection{Stoichiometric Capacity}
From the net reaction (\cref{eq:scrub_primary}), 1 mole of Ca(OH)\textsubscript{2} consumes 1 mole of CO\textsubscript{2}. With the molecular masses $M_{\mathrm{Ca(OH)_2}} = \SI{74.09}{\gram\per\mole}$ and $M_{\mathrm{CO_2}} = \SI{44.01}{\gram\per\mole}$, the scrubbing capacity depends on the available Ca(OH)\textsubscript{2} in the as-packed canister. Soda lime as packed contains $\sim$15--20\% water by mass (required for the aqueous film that enables scrubbing); the dry mass fraction is therefore $\sim$0.80--0.85 of the total, and Ca(OH)\textsubscript{2} constitutes $\sim$75--80\% of the dry mass:
\begin{equation}
\label{eq:scrub_capacity}
m_{\mathrm{CO_2,max}} = \frac{f_{\mathrm{dry}} \times f_{\mathrm{Ca(OH)_2}} \times m_{\mathrm{soda\,lime}}}{M_{\mathrm{Ca(OH)_2}}} \times M_{\mathrm{CO_2}} = \frac{0.82 \times 0.77 \times \SI{1000}{\gram}}{\SI{74.09}{\gram\per\mole}} \times \SI{44.01}{\gram\per\mole} \approx \SI{375}{\gram}
\end{equation}
where $f_{\mathrm{dry}} \approx 0.82$ is the dry mass fraction (assuming 18\% water, mid-range) and $f_{\mathrm{Ca(OH)_2}} \approx 0.77$ is the Ca(OH)\textsubscript{2} fraction of dry mass. The canister can absorb a maximum of $\sim$\SI{375}{\gram} of CO\textsubscript{2}. At a moderate metabolic CO\textsubscript{2} production rate of approximately \SI{200}{\milli\liter\per\minute} (STP), corresponding to $\dot{m}_{\mathrm{CO_2}} \approx \SI{0.39}{\gram\per\minute}$, the scrubber lifetime is:
\begin{equation}
T_{\mathrm{scrubber}} \approx \frac{375}{0.39} \approx \SI{962}{\minute} \approx \SI{16}{\hour} \quad\text{(at rest)}
\end{equation}
Under heavy exertion, $\dot{V}_{\mathrm{CO_2}}$ can rise to \SIrange{2.0}{3.0}{\liter\per\minute}, reducing scrubber life to approximately \SIrange{1.0}{1.6}{\hour}.
\subsubsection{Thermal Management}
\label{sec:thermal}
From \cref{eq:enthalpy}, the heat generation rate in the scrubber is:
\begin{equation}
\label{eq:heat_scrub}
\dot{Q}_{\mathrm{scrub}}(t) = |\Delta H_{\mathrm{rxn}}^\circ| \times \dot{n}_{\mathrm{CO_2,scrubbed}}(t)
\end{equation}
where $\dot{n}_{\mathrm{CO_2,scrubbed}}$ is the molar scrubbing rate. At heavy exertion ($\dot{n}_{\mathrm{CO_2}} \approx \SI{0.068}{\mole\per\minute}$):
\begin{equation}
\dot{Q}_{\mathrm{scrub}} \approx \frac{113.1 \times 0.068}{60} \,\si{\kilo\joule\per\second} \approx \SI{128}{\watt}
\end{equation}
Combined with the desiccant heat of adsorption ($\dot{Q}_{\mathrm{ads}} \approx \SI{179}{\watt}$ at peak, \cref{eq:heat_ads}), the total internal heat generation from the treatment train reaches up to $\sim$\SI{307}{\watt} at heavy exertion.

\medskip
\noindent\textbf{Why no external heat exchanger is used:}
In a structural fire, the external environment is frequently hotter than the breathing loop. Any attempt to reject heat to the suit shell can reverse sign and \emph{add} heat to the loop when $T_{\mathrm{ext}} > T_{\mathrm{loop}}$. For this reason the design does not rely on a shell-coupled heat exchanger for cooling; thermal safety is instead managed by (i) limiting internal heat generation via control (fan speed, scrubber bypass), and (ii) controlled outward venting strategies described below.

\medskip
\noindent\textbf{Supplementary evaporative venting:}
\label{sec:evap_vent}
For extended missions, the system is supplemented by \emph{controlled evaporative venting}: a small amount of water (condensed from the desiccant or from a separate reservoir) is released through a one-way valve to the exterior, where it evaporates, carrying away $\sim$\SI{2440}{\kilo\joule\per\kilogram}. This breaks the ``fully closed'' thermal constraint at a modest water cost ($\sim$\SI{6}{\gram\per\minute} to reject \SI{250}{\watt}). 
Design mitigations include:
\begin{itemize}[nosep]
    \item Granular soda lime bed with controlled void fraction ($\varepsilon \approx 0.35$--$0.45$) to allow convective heat removal by the circulating gas.
    \item Thermal fuse: if the scrubber bed thermocouple registers $T_{\mathrm{bed}} > \SI{80}{\celsius}$, the controller automatically increases scrubber bypass fraction $\phi_{\mathrm{bypass}}$ and fan speed, trading temporarily elevated CO\textsubscript{2} for thermal safety.
\end{itemize}
The transient temperature of the scrubber bed can be modeled using an energy balance:
\begin{equation}
\label{eq:bed_temp}
(\rho c_p)_{\mathrm{bed}} V_{\mathrm{bed}} \frac{dT_{\mathrm{bed}}}{dt} = \dot{Q}_{\mathrm{scrub}}(t) - \dot{m}_{\mathrm{air}} c_{p,\mathrm{air}} (T_{\mathrm{bed}} - T_{\mathrm{air,in}}) - U A_{\mathrm{wall}} (T_{\mathrm{bed}} - T_{\mathrm{wall}})
\end{equation}
where $(\rho c_p)_{\mathrm{bed}}$ is the effective volumetric heat capacity of the packed bed, $U$ is the overall heat transfer coefficient to the canister wall.
\section{Humidity Management}
\label{sec:humidity}
\subsection{Source of Moisture and Interaction with Scrubber Placement}
In a closed-circuit system, water vapor has two distinct sources, both of which must be managed by the desiccant:

\begin{enumerate}[label=(\roman*)]
    \item \textbf{Metabolic moisture:} The firefighter's exhaled breath and insensible perspiration. A resting adult exhales approximately \SIrange{200}{400}{\milli\liter} of liquid-equivalent water per day via respiration; under heavy exertion and thermal stress, this can increase five-fold or more, reaching \SIrange{1.5}{3.0}{\gram\per\minute}.
    \item \textbf{Reaction-generated moisture:} The soda lime scrubbing reaction (\cref{eq:scrub_primary}) produces one mole of H\textsubscript{2}O per mole of CO\textsubscript{2} absorbed. At heavy exertion, this contributes an additional $\sim$\SI{1.2}{\gram\per\minute} (\cref{eq:scrub_water}), representing \SIrange{30}{45}{\percent} of the total moisture load.
\end{enumerate}

As discussed in \cref{sec:airflow_pathway}, the silica gel desiccant is positioned \emph{downstream} of the scrubber so that it intercepts both sources in a single pass. This topology also preserves the moist gas environment required for efficient scrubber operation. The system targets a relative humidity (RH) below \SI{60}{\percent} at the suit breathing zone, measured by the in-suit capacitive RH sensor downstream of the desiccant stage.
Excessive humidity causes visor fogging (onset at RH~$> 80\%$ on cool visor surfaces), skin maceration, and reduced evaporative cooling efficiency---further elevating the firefighter's core temperature in an already thermally hostile environment.
\subsection{Silica Gel Adsorption: Physical Chemistry}
Silica gel ($\mathrm{SiO_2 \cdot nH_2O}$) is an amorphous, highly porous form of silicon dioxide. Its internal surface area---typically \SIrange{600}{800}{\meter\squared\per\gram}---provides abundant sites for physical adsorption (physisorption) of water molecules via hydrogen bonding with surface silanol ($\mathrm{Si\text{-}OH}$) groups.
\subsubsection{Adsorption Isotherm}
The equilibrium moisture uptake of silica gel as a function of relative humidity exhibits strong multilayer adsorption and capillary condensation behavior, particularly above $\sim$40\% RH where the humidity control constraint operates. The simple Langmuir isotherm (monolayer, asymptotic saturation) systematically underpredicts uptake in this regime. We therefore use the \emph{Guggenheim--Anderson--de Boer (GAB)} isotherm, a three-parameter extension of BET theory that is standard in adsorption engineering for water/silica gel systems:
\begin{equation}
\label{eq:gab}
q_e = \frac{q_m \, C_G \, K_G \, a_w}{(1 - K_G \, a_w)(1 - K_G \, a_w + C_G \, K_G \, a_w)}
\end{equation}
where $q_e$ is the equilibrium loading (kg water per kg dry silica), $a_w = \mathrm{RH}/100$ is the water activity, $q_m$ is the monolayer capacity, $C_G$ is the Guggenheim constant (related to the enthalpy difference between monolayer and multilayer adsorption), and $K_G$ is a multilayer correction factor. For Type~A silica gel at \SI{25}{\celsius}, representative parameters are $q_m \approx 0.10$, $C_G \approx 40$, $K_G \approx 0.85$, yielding $q_e \approx 0.30$--$0.35$ at 80\% RH---consistent with published data. All three parameters are temperature-dependent (decreasing $q_e$ with increasing temperature at constant RH), which is captured by Arrhenius-type expressions fitted to manufacturer data.
\subsubsection{Adsorption Dynamics and Linear Driving Force Model}
The rate of water uptake is governed by intraparticle diffusion through the pore network. Using the linear driving force (LDF) approximation:
\begin{equation}
\label{eq:ldf}
\frac{d\bar{q}(t)}{dt} = k_{\mathrm{LDF}} \bigl[ q_e\bigl(\mathrm{RH}_{\mathrm{in}}(t)\bigr) - \bar{q}(t) \bigr]
\end{equation}
where $\bar{q}(t)$ is the average loading on the gel at time $t$ and $k_{\mathrm{LDF}}$ is the LDF mass transfer coefficient, which depends on effective pore diffusivity $D_e$, bead radius $R_p$, and the approximation $k_{\mathrm{LDF}} \approx 15 D_e / R_p^2$.
\subsubsection{Mass Balance for the Packed Bed}
For the silica gel canister, the macroscopic water mass balance is:
\begin{equation}
\label{eq:silica_mass}
\frac{dM_{\mathrm{water}}(t)}{dt} = \dot{m}_{\mathrm{air}} \bigl[ Y_{\mathrm{in}}(t) - Y_{\mathrm{out}}(t) \bigr]
\end{equation}
where $Y_{\mathrm{in}}(t)$ and $Y_{\mathrm{out}}(t)$ are the humidity ratios (kg water per kg dry air) at the canister inlet and outlet, respectively, and $\dot{m}_{\mathrm{air}}$ is the dry-air mass flow rate. The total silica canister mass evolves as:
\begin{equation}
M_{\mathrm{silica}}(t) = M_{\mathrm{silica,dry}} + M_{\mathrm{water}}(t), \quad 0 \le M_{\mathrm{water}}(t) \le M_{\mathrm{water,max}}
\end{equation}
As $M_{\mathrm{water}}(t) \to M_{\mathrm{water,max}} \approx \SI{0.35}{\kilogram}$ (for \SI{1}{\kilogram} of dry silica gel), $Y_{\mathrm{out}} \to Y_{\mathrm{in}}$ and the desiccant is effectively saturated.
Expressing the driving force in terms of partial pressures:
\begin{equation}
\frac{dM_{\mathrm{water}}(t)}{dt} = \dot{m}_{\mathrm{air}} \, \alpha \bigl[ p_{\mathrm{H_2O,in}}(t) - p_{\mathrm{H_2O,out}}(t) \bigr]
\end{equation}
where $\alpha$ is a lumped proportionality constant accounting for the psychrometric relationship between humidity ratio and partial pressure at the system's operating temperature and total pressure.
\subsubsection{Heat of Adsorption and the Total Thermal Budget}
Water adsorption onto silica gel is exothermic. The enthalpy of adsorption $\Delta H_{\mathrm{ads}}$ includes the latent heat of condensation of water vapor ($\Delta H_{\mathrm{vap}} \approx \SI{2440}{\kilo\joule\per\kilogram}$ at \SI{35}{\celsius}) plus the excess heat of surface binding ($\Delta H_{\mathrm{excess}} \approx \SIrange{100}{200}{\kilo\joule\per\kilogram}$ for Type~A silica gel, depending on loading). The total enthalpy of adsorption is therefore:
\begin{equation}
\Delta H_{\mathrm{ads}} \approx \SIrange{2500}{2600}{\kilo\joule\per\kilogram\text{ H}_2\text{O}}
\end{equation}
The heat generation rate in the desiccant bed is:
\begin{equation}
\label{eq:heat_ads_rate}
\dot{Q}_{\mathrm{ads}}(t) = \Delta H_{\mathrm{ads}} \times \frac{dM_{\mathrm{water}}(t)}{dt}
\end{equation}
At peak moisture load (\SI{4.2}{\gram\per\minute}, see \cref{sec:moisture_budget}), this yields $\dot{Q}_{\mathrm{ads}} \approx \SI{179}{\watt}$---a heat load comparable to the scrubber's output (\SI{128}{\watt} at heavy exertion). The combined thermal output of both packed beds is therefore up to $\sim$\textbf{\SI{307}{\watt}} (\cref{sec:thermal}). The desiccant bed temperature can be modeled analogously to \cref{eq:bed_temp}:
\begin{equation}
(\rho c_p)_{\mathrm{silica}} V_{\mathrm{silica}} \frac{dT_{\mathrm{silica}}}{dt} = \dot{Q}_{\mathrm{ads}}(t) - \dot{m}_{\mathrm{air}} c_{p,\mathrm{air}} (T_{\mathrm{silica}} - T_{\mathrm{air,in}}) - U_{\mathrm{silica}} A_{\mathrm{wall,silica}} (T_{\mathrm{silica}} - T_{\mathrm{wall,silica}})
\end{equation}
\section{Oxygen Replenishment}
\label{sec:oxygen}
\subsection{Metabolic Basis}
Aerobic metabolism is summarized by the oxidation of glucose:
\begin{equation}
\label{eq:glucose}
\mathrm{C_6H_{12}O_6(aq) + 6\,O_2(g) \longrightarrow 6\,CO_2(g) + 6\,H_2O(l)} \qquad \Delta H^\circ = \SI{-2803}{\kilo\joule\per\mole}
\end{equation}
The respiratory exchange ratio (RER), defined as $R = \dot{V}_{\mathrm{CO_2}} / \dot{V}_{\mathrm{O_2}}$, ranges from $\sim 0.7$ (pure fat oxidation) to $1.0$ (pure carbohydrate oxidation) and exceeds 1.0 above the respiratory compensation point, when excess CO\textsubscript{2} is produced from bicarbonate buffering of lactic acid. At moderate exertion, $R \approx 0.85$.
\subsection{Oxygen Consumption Rates}
A firefighter's metabolic O\textsubscript{2} consumption rate, $\dot{V}_{\mathrm{O_2}}$, depends strongly on work rate $W(t)$:
\begin{equation}
\label{eq:vo2}
\dot{V}_{\mathrm{O_2}}(t) = \dot{V}_{\mathrm{O_2,rest}} + \gamma \, W(t) + \beta \, [W(t)]^2
\end{equation}
where $\dot{V}_{\mathrm{O_2,rest}} \approx \SI{0.25}{\liter\per\minute}$ (STP), $\gamma$ and $\beta$ are subject-specific coefficients capturing the additional oxygen cost at high exertion (the $\dot{V}_{\mathrm{O_2}}$ slow component, reflecting increased reliance on fast-twitch motor units above the lactate threshold), and $W(t)$ is the instantaneous metabolic work rate (\si{\watt}). Peak $\dot{V}_{\mathrm{O_2}}$ for an elite firefighter can reach \SIrange{3.0}{4.0}{\liter\per\minute}, corresponding to a mass consumption rate of:
\begin{equation}
\dot{m}_{\mathrm{O_2}} = \dot{V}_{\mathrm{O_2}} \times \frac{M_{\mathrm{O_2}}}{V_m} = \dot{V}_{\mathrm{O_2}} \times \frac{32.00}{22.414} \approx 1.43 \, \dot{V}_{\mathrm{O_2}} \quad [\si{\gram\per\minute}]
\end{equation}
\subsection{Oxygen Tank Endurance}
The finite oxygen supply of \SI{3.0}{\kilogram} constrains mission duration:
\begin{equation}
\label{eq:o2_constraint}
\int_0^{T_{\max}} \dot{m}_{\mathrm{O_2,inject}}(t)\,dt \le \SI{3000}{\gram}
\end{equation}
The O\textsubscript{2} injection rate must compensate for both metabolic consumption and the gas lost through exhaust valve venting. At peak metabolic consumption alone (\SI{4.0}{\liter\per\minute} STP), $\dot{m}_{\mathrm{O_2}} \approx \SI{5.7}{\gram\per\minute}$, giving a \emph{metabolic-only} endurance of $3000/5.7 \approx \SI{526}{\minute}$. However, the exhaust valve venting creates an additional O\textsubscript{2} demand: each vent event exhausts gas at the current loop composition, which must be replaced with pure O\textsubscript{2}. The effective endurance depends on the venting rate, which in turn depends on thermal transients, body movement, and the pressure control strategy.
At moderate sustained exertion ($\dot{V}_{\mathrm{O_2}} \approx \SI{2.0}{\liter\per\minute}$) with a typical vent loss of $\sim$\SI{0.5}{\liter\per\minute} (equivalent), the combined O\textsubscript{2} consumption is $\dot{m}_{\mathrm{O_2}} \approx \SI{3.6}{\gram\per\minute}$, yielding an endurance of $\sim$\SI{830}{\minute} ($\sim$14 hours). At heavy sustained exertion with frequent venting, endurance drops to $\sim$\SIrange{3}{5}{\hour}---still substantially exceeding the 30-minute limit of open-circuit SCBA. This regime is precisely where intelligent control adds value: the MPC optimizes the tradeoff between O\textsubscript{2} conservation, vent frequency management, and physiological safety.
\subsection{Oxygen Toxicity Considerations}
In a pure-O\textsubscript{2} replenishment system, the inspired partial pressure of O\textsubscript{2} ($P_iO_2$) must be carefully controlled. At sea-level ambient pressure ($\sim$\SI{1}{\atm}), breathing \SI{100}{\percent} O\textsubscript{2} yields $P_iO_2 \approx \SI{1.0}{\atm}$. Prolonged exposure above $P_iO_2 > \SI{0.5}{\atm}$ risks \emph{pulmonary oxygen toxicity} (Lorrain Smith effect) \cite{clark_lambertsen_o2}, while $P_iO_2 > \SI{1.6}{\atm}$ can precipitate \emph{central nervous system (CNS) oxygen toxicity} with seizures \cite{oxygentoxicity, mixedgas, lungdiffusingcapacity, pulmonarygas}.
The control system must therefore maintain $P_iO_2$ within a safe band:
\begin{equation}
\SI{0.19}{\atm} \le P_iO_2(t) \le \SI{0.50}{\atm}
\end{equation}
corresponding to an inspired O\textsubscript{2} fraction of approximately \SIrange{19}{50}{\percent} at \SI{1}{\atm}.
\subsection{Oxygen Enrichment as a Fire Hazard}
\label{sec:o2_fire}
The paper's O\textsubscript{2} toxicity constraint ($P_iO_2 \le \SI{0.50}{\atm}$, or $\sim$50\% O\textsubscript{2}) is insufficient as the sole upper bound on oxygen fraction. Per NFPA standards and oxygen-system safety engineering (NASA, ASTM), an atmosphere above $\sim$23.5\% O\textsubscript{2} by volume is classified as \emph{oxygen-enriched}, with significantly increased fire risk: materials that are self-extinguishing in air may burn vigorously, ignition energies decrease, and flame propagation rates increase. In a firefighting suit operating \emph{in a fire environment}, this risk is compounded.
The O\textsubscript{2} fraction in the breathing loop must therefore be constrained by \emph{fire safety} as well as toxicity:
\begin{equation}
\label{eq:o2_fire}
x_{\mathrm{O_2}}(t) \le x_{\mathrm{O_2,fire}} = 0.235
\end{equation}
where $x_{\mathrm{O_2,fire}} = 0.235$ is the oxygen-enriched-atmosphere
threshold used here for fire safety, based on OSHA guidance \cite{osha_o2}.
This is a substantially tighter constraint than the toxicity limit of 0.50 and fundamentally shapes the control problem. The fire-safety limit is the \emph{binding} upper constraint under normal operation; the toxicity limit serves as a hard backup for degraded-mode operation where maintaining 23.5\% is no longer feasible.
\begin{remark}[Implications for system architecture]
The fire-safety O\textsubscript{2} limit strongly motivates confining the O\textsubscript{2}-enriched breathing gas to a \emph{small internal breathing loop} (mask, hoses, counter-lung, treatment train) rather than flooding the entire suit interior volume. In future design iterations, a separate low-O\textsubscript{2} gas (filtered air or N\textsubscript{2}-enriched mix) could pressurize the suit shell, with the breathing loop isolated by a mask/mouthpiece. This paper analyzes the simpler single-atmosphere architecture and applies the 23.5\% constraint to the full suit volume.
\end{remark}
\section{Airflow Physics and Circulation}
\label{sec:airflow}
\subsection{Positive Pressure Maintenance and Gas Inventory Dynamics}
\label{sec:gas_dynamics}
\subsubsection{Correct Gas-Phase Molar Bookkeeping}
In a closed-circuit breathing system with chemical CO\textsubscript{2} scrubbing, the gas-phase molar inventory evolves as follows. Consider one metabolic--scrubbing cycle:
\begin{enumerate}[nosep]
    \item The firefighter consumes 1~mol of O\textsubscript{2} from the gas phase (uptake into blood). Gas moles: $-1$.
    \item The firefighter exhales $R$~mol of CO\textsubscript{2} (where $R \approx 0.85$ is the respiratory exchange ratio) \emph{into} the gas phase. Gas moles: $-1 + R$.
    \item The scrubber removes the $R$~mol of CO\textsubscript{2} from the gas phase, converting it to solid CaCO\textsubscript{3}. Gas moles: $-1 + R - R = -1$.
    \item The O\textsubscript{2} injection system adds 1~mol of O\textsubscript{2}. Gas moles: $-1 + 1 = 0$.
\end{enumerate}
The net change in gas-phase moles per cycle is \textbf{zero}---the CO\textsubscript{2} produced by metabolism and removed by scrubbing exactly cancel in the molar bookkeeping, and the consumed O\textsubscript{2} is replaced by injection. There is no inherent ``molar sink'' from the scrubbing chemistry in a sealed system with 1:1 O\textsubscript{2} replacement.
\begin{remark}[Common error in molar-sink analyses]
A frequently encountered error is to treat the scrubber as removing CO\textsubscript{2} from the loop's \emph{initial} inventory rather than from the CO\textsubscript{2} just produced by metabolism. This double-counts the CO\textsubscript{2}: the body adds $R$~mol to the gas phase and the scrubber immediately removes $R$~mol from the gas phase. The net effect on gas-phase moles is zero from the CO\textsubscript{2} pathway. The only gas-phase deficit is the $-1$~mol O\textsubscript{2} consumed, which is exactly compensated by injection.
\end{remark}
\subsubsection{Exhaust Valve Venting: The Real Driver of Gas Inventory Change}
In the Galactic Bioware semi-closed suit, the \emph{exhaust valves} are the primary mechanism that changes the gas-phase inventory. When internal pressure exceeds the cracking pressure $P_{\mathrm{crack}}$, gas vents outward. The valve behaves as a compressible orifice; in the subsonic regime (applicable for the small $\Delta P$ involved), the mass flow rate follows:
\begin{equation}
\label{eq:vent_rate}
\dot{n}_{\mathrm{vent}}(t) = \frac{C_d A_v}{\bar{M}} \sqrt{2 \rho_s \, \max\!\bigl(0, \; P_s(t) - P_{\mathrm{crack}}\bigr)}
\end{equation}
where $C_d$ is the valve discharge coefficient, $A_v$ is the effective valve orifice area, $\bar{M}$ is the mean molar mass of the loop gas, and $\rho_s$ is the gas density at suit conditions. The $\sqrt{\Delta P}$ dependence---rather than a linear relationship---is the standard orifice flow law and is important because the O\textsubscript{2} enrichment analysis (\cref{sec:o2_enrichment}) is directly sensitive to the vent rate functional form. The vented gas has the \emph{current loop composition}: a mixture of N\textsubscript{2}, O\textsubscript{2}, residual CO\textsubscript{2}, and water vapor. This depletes all species proportionally to their mole fractions.
Pressure rises (triggering venting) occur due to:
\begin{itemize}[nosep]
    \item Body movement compressing the suit volume (bending, crouching, impact)
    \item Thermal expansion of the gas from external heating
    \item O\textsubscript{2} injection temporarily exceeding metabolic consumption
    \item Tidal exhalation peaks
\end{itemize}
\subsubsection{Oxygen Enrichment from Vent Compensation}
\label{sec:o2_enrichment}
Because vented gas contains the current N\textsubscript{2}/O\textsubscript{2} mixture but is replaced with \emph{pure} O\textsubscript{2}, each vent--refill cycle increases the O\textsubscript{2} mole fraction. The enrichment dynamic is:
\begin{equation}
\label{eq:o2_enrichment_exact}
\frac{dx_{\mathrm{O_2}}}{dt}
=
\frac{(1-x_{\mathrm{O_2}})\,\bigl(\dot{n}_{\mathrm{O_2,inject}} - \dot{n}_{\mathrm{O_2,consumed}}\bigr)}{n_{\mathrm{total}}}
\end{equation}
Under the pressure-holding approximation
\[
\dot{n}_{\mathrm{O_2,inject}} \approx \dot{n}_{\mathrm{O_2,consumed}} + \dot{n}_{\mathrm{vent}},
\]
the enrichment dynamics reduce to:
\begin{equation}
\label{eq:o2_enrichment_simple}
\frac{dx_{\mathrm{O_2}}}{dt}
\approx
\frac{\dot{n}_{\mathrm{vent}}}{n_{\mathrm{total}}}(1-x_{\mathrm{O_2}})
\end{equation}
For a reactive packed bed, the appropriate sizing relation is obtained from
a plug-flow absorber formulation:
\begin{equation}
\frac{d\dot{n}_{\mathrm{CO_2}}}{dz}
=
- k_{\mathrm{ov}} a_s A_c
\left(p_{\mathrm{CO_2}} - p_{\mathrm{CO_2}}^*\right)
\end{equation}
Integrating along the bed length yields the required number of transfer units (NTU):
\begin{equation}
\ln\!\left(
\frac{p_{\mathrm{CO_2,in}} - p_{\mathrm{CO_2}}^*}
{p_{\mathrm{CO_2,out}} - p_{\mathrm{CO_2}}^*}
\right)
=
\mathrm{NTU}
=
\frac{k_{\mathrm{ov}} a_s V_{\mathrm{bed}}}{\dot{V}_{\mathrm{circ}}}
\end{equation}
The circulation rate is therefore determined by required NTU and mass-transfer
kinetics rather than by simple dilution assumptions.
The enrichment rate is proportional to the \emph{vent rate}, not to the metabolic or scrubbing rate. Starting from air ($x_{\mathrm{O_2}} = 0.21$), the O\textsubscript{2} fraction rises gradually as N\textsubscript{2} is diluted. The time to reach the fire-safety limit of 23.5\% depends strongly on the venting frequency, which is controlled by body movement patterns, thermal transients, and the pressure control strategy. For a loop containing $n_{\mathrm{total}} \approx 4$~mol ($\sim$\SIrange{95}{100}{\liter} at \SI{1}{\atm}, \SI{35}{\celsius}) and an upper-bound average vent rate of $\sim\SI{0.05}{\mole\per\minute}$ (equivalent to $\sim$\SI{1.1}{\liter\per\minute} STP):
\begin{equation}
\frac{dx_{\mathrm{O_2}}}{dt}\bigg|_{x_{\mathrm{O_2}}=0.21} \approx \frac{0.05}{4}(1 - 0.21) \approx 0.010\;\text{min}^{-1}
\end{equation}
The fire-safety limit of $x_{\mathrm{O_2}} = 0.235$ would be reached in approximately $0.025 / 0.010 \approx 2.5$~minutes at this sustained vent rate.  This should be interpreted as a bounding worst-case: $\dot{n}_{\mathrm{vent}} \sim \SI{0.05}{\mole\per\minute}$ corresponds to $\sim$\SI{1.1}{\liter\per\minute} STP-equivalent, which is substantially higher than typical leak/vent rates reported for positive-pressure suits at \SIrange{2}{5}{\milli\bar} overpressure. If the effective vent rate is closer to \SIrange{0.1}{0.3}{\liter\per\minute} equivalent, the enrichment timescale increases proportionally to $\sim$\SIrange{8}{25}{\minute} for the same loop volume. This is an extremely tight timeline that renders the single-atmosphere architecture operationally marginal for anything beyond very short deployments with minimal body movement (and hence minimal venting). \textbf{This result should be understood as a bounding analysis}: the single-atmosphere design represents the worst case for O\textsubscript{2} enrichment because the entire suit volume ($\sim$\SI{100}{\liter}) participates in the vent--refill cycle. Typical positive-pressure suits exhibit leak/vent rates closer to \SIrange{0.1}{0.3}{\liter\per\minute} at \SIrange{2}{5}{\milli\bar} overpressure. The \SI{0.05}{\mole\per\minute} value therefore represents a stress-test scenario rather than a nominal operating condition. A separated architecture (breathing loop confined to mask/counter-lung/treatment train at $\sim$\SI{10}{\liter}, with suit pressurization via an inert or filtered-air source) would reduce the enrichment rate by roughly an order of magnitude. The AI controller's ability to manage the enrichment--pressure tradeoff is therefore necessary but likely insufficient on its own; the separated architecture discussed in the future-work section is the viable path to fielded hardware. Nevertheless, the control problem structure and the MPC formulation remain valid regardless of the loop volume, so we analyze the harder single-atmosphere case:
\begin{itemize}[nosep]
    \item \textbf{High positive pressure margin}: better suit integrity protection but more frequent venting, faster O\textsubscript{2} enrichment.
    \item \textbf{Low positive pressure margin}: reduced venting and slower enrichment but increased risk of ambient gas infiltration through suit imperfections.
\end{itemize}
The MPC optimizes this tradeoff in real time.
\subsubsection{Counter-Lung Dynamics and Suit Pressure}
The counter-lung accommodates transient volume fluctuations. Rather than treating suit pressure $P_s$ as an independent dynamic state---which would create a differential-algebraic inconsistency with the approximately isobaric regime enforced by the compliant counter-lung---we model $P_s$ algebraically from the counter-lung's elastic restoring force:
\begin{equation}
\label{eq:ps_algebraic}
P_s(t) = P_a(t) + k_{\mathrm{CL}} \bigl(V_{\mathrm{CL}}(t) - V_{\mathrm{CL,0}}\bigr)
\end{equation}
where $k_{\mathrm{CL}}$ is the counter-lung stiffness (\si{\pascal\per\liter}), a small value for a compliant bellows (typically \SIrange{50}{200}{\pascal\per\liter}), and $V_{\mathrm{CL,0}}$ is the neutral volume. The exhaust valve opens when $P_s > P_{\mathrm{crack}}$, i.e., when $V_{\mathrm{CL}}$ exceeds the threshold $V_{\mathrm{CL,0}} + (P_{\mathrm{crack}} - P_a)/k_{\mathrm{CL}}$. This couples the vent dynamics naturally to the counter-lung state without requiring a separate pressure ODE.
The counter-lung volume evolves according to:
\begin{equation}
\label{eq:counterlung_dynamics}
\frac{dV_{\mathrm{CL}}}{dt}
=
\underbrace{\frac{R_g T_{\mathrm{suit}}}{P_s}\bigl(\dot{n}_{\mathrm{O_2,inject}} - \dot{n}_{\mathrm{O_2,consumed}} - \dot{n}_{\mathrm{vent}}\bigr)}_{\text{net molar change}}
+
\underbrace{\frac{V_{\mathrm{gas}}(t)}{T_{\mathrm{suit}}(t)}\,\frac{dT_{\mathrm{suit}}}{dt}}_{\text{thermal expansion}}
+
\dot{V}_{\mathrm{breath}}(t)
\end{equation}
The first term captures the net molar balance: O\textsubscript{2} injected minus O\textsubscript{2} consumed minus gas vented. When the controller tracks metabolic consumption precisely and venting is minimal, this term is near zero and the counter-lung volume remains stable. The thermal expansion term accounts for gas expansion/contraction with temperature. $\dot{V}_{\mathrm{breath}}(t)$ is the oscillatory tidal breathing flow term.

\medskip
\noindent\textbf{Thermal expansion compensation in the EKF:}
The raw counter-lung position measurement $V_{\mathrm{CL}}^{\mathrm{meas}}(t)$ conflates molar changes with thermal expansion. In a catastrophic fire scenario where external temperatures escalate rapidly, the thermal expansion term can inflate the counter-lung even as gas is being lost through venting. The EKF incorporates the thermal expansion term directly in its process model, using the in-suit temperature measurements ($T_{\mathrm{suit,bz}}$ and $T_{\mathrm{suit,torso}}$) to separate thermal effects from molar changes. This cross-modal consistency check---volume sensor versus temperature sensors versus gas composition sensors---is a key advantage of the multi-sensor EKF architecture.
\subsection{Internal Circulation: Fan Dynamics}
The fans do not exchange mass with the environment; they circulate gas through the treatment train. The volumetric flow rate through the closed loop is:
\begin{equation}
Q_{\mathrm{circ}} = \frac{\Delta P_{\mathrm{fan}}}{R_{\mathrm{sys}}}
\end{equation}
where $\Delta P_{\mathrm{fan}}$ is the fan's developed pressure and $R_{\mathrm{sys}}$ is the total flow resistance. The gas passes sequentially through the soda lime scrubber, and the silica gel canister (plus interconnecting tubing and the suit interior). For a packed bed of spherical particles, the pressure drop per unit length is given by the Ergun equation \cite{ergun}:
\begin{equation}
\label{eq:ergun}
\frac{\Delta P}{L} = \frac{150 \mu (1-\varepsilon)^2}{\varepsilon^3 d_p^2} v_s + \frac{1.75 \rho (1-\varepsilon)}{\varepsilon^3 d_p} v_s^2
\end{equation}
where $\mu$ is gas dynamic viscosity, $\varepsilon$ is bed void fraction, $d_p$ is particle diameter, $v_s$ is superficial velocity, $L$ is bed length, and $\rho$ is gas density.
\subsubsection{Time-Varying Void Fraction from Solid and Liquid Volume Expansion}
The Ergun equation is conventionally applied with a constant $\varepsilon$, but the soda lime scrubber bed undergoes a volume change as the reaction proceeds. The reactant Ca(OH)\textsubscript{2} has a molar volume of $\bar{V}_{\mathrm{Ca(OH)_2}} \approx \SI{33.0}{\centi\meter\cubed\per\mole}$, while the product CaCO\textsubscript{3} (calcite) has $\bar{V}_{\mathrm{CaCO_3}} \approx \SI{36.9}{\centi\meter\cubed\per\mole}$. Crucially, the reaction also produces 1~mol of liquid H\textsubscript{2}O per mol of CO\textsubscript{2} (\cref{eq:scrub_primary}), with molar volume $\bar{V}_{\mathrm{H_2O(l)}} \approx \SI{18.0}{\centi\meter\cubed\per\mole}$. The paper explicitly requires this water to remain as an aqueous surface film on the granules for scrubbing to function; it therefore occupies void space within the bed.
Not all reaction water remains in the bed: a fraction evaporates into the gas stream and is carried downstream to the desiccant (this is the moisture load already accounted for in \cref{sec:airflow_pathway}). We introduce a water retention fraction $\chi_w \in [0,1]$, where $\chi_w = 1$ means all reaction water remains in the bed and $\chi_w = 0$ means it all evaporates. The effective volume displacing voids per mole reacted is then:
\begin{equation}
\bar{V}_{\mathrm{products}}(\chi_w) = \bar{V}_{\mathrm{CaCO_3}} + \chi_w \, \bar{V}_{\mathrm{H_2O(l)}} = 36.9 + 18.0\,\chi_w \quad [\si{\centi\meter\cubed\per\mole}]
\end{equation}
Let $\xi(t) \in [0,1]$ be the fractional conversion of Ca(OH)\textsubscript{2}. The combined solid-plus-liquid volume in the canister evolves as:
\begin{equation}
V_{\mathrm{solid+liquid}}(t) = V_{\mathrm{solid,0}} \bigl[1 + \xi(t)(\sigma(\chi_w) - 1)\bigr]
\end{equation}
where the effective swelling ratio is:
\begin{equation}
\sigma(\chi_w) = \frac{\bar{V}_{\mathrm{CaCO_3}} + \chi_w \, \bar{V}_{\mathrm{H_2O(l)}}}{\bar{V}_{\mathrm{Ca(OH)_2}}} = \frac{36.9 + 18.0\,\chi_w}{33.0}
\end{equation}
At $\chi_w = 0$ (all water evaporated): $\sigma = 1.12$, recovering the solid-only analysis. At $\chi_w = 0.5$ (half retained): $\sigma \approx 1.39$. At $\chi_w = 1$ (all water retained): $\sigma \approx 1.66$. The void fraction in the fixed-volume canister then decreases:
\begin{equation}
\label{eq:epsilon_t}
\varepsilon(t) = 1 - (1 - \varepsilon_0)\bigl[1 + \xi(t)(\sigma(\chi_w) - 1)\bigr]
\end{equation}
where $\varepsilon_0$ is the initial void fraction. The following table summarizes the impact:
\begin{center}
\begin{tabular}{lSSS}
\toprule
& {$\chi_w = 0$} & {$\chi_w = 0.5$} & {$\chi_w = 1$} \\
\midrule
$\sigma$ & 1.12 & 1.39 & 1.66 \\
$\varepsilon(\xi{=}1)$ for $\varepsilon_0 = 0.40$ & 0.33 & 0.17 & 0.00 \\
Approx.\ flow resistance increase\textsuperscript{\dag} & {$2.2\times$} & {$25\times$} & {$\to \infty$} \\
\bottomrule
\end{tabular}
\end{center}
\noindent\textsuperscript{\dag}Estimated from the Ergun equation's dominant viscous term, which scales as $(1-\varepsilon)^2/\varepsilon^3$. The ratio at $\varepsilon = 0.33$ vs.\ $\varepsilon_0 = 0.40$ is $(0.67^2/0.33^3)/(0.60^2/0.40^3) = 12.5/5.6 \approx 2.2\times$; at $\varepsilon \approx 0.17$ (rounded from 0.166) the ratio is
$(0.83^2/0.17^3)/(0.60^2/0.40^3) \approx 25\times$; using the
unrounded value $\varepsilon = 0.166$ gives $\approx 27\times$.
Both are reported as $\sim\!25\times$ given the larger uncertainty
in $\chi_w$.
The $\chi_w = 1$ case (complete bed blockage) is physically unrealistic because the bed would choke and force gas through bypass channels long before complete conversion. In practice, the gas flow through the bed actively evaporates surface water, so $\chi_w$ is itself a function of flow rate, temperature, and humidity: at high circulation rates and low inlet humidity, $\chi_w$ is driven toward zero. The Ergun equation's $\varepsilon^3$ denominator amplifies even modest void-fraction reductions into large flow resistance increases, making this a critical coupling: the state-space model must track $\xi(t)$, and the sensitivity to $\chi_w$ should be included in the MPC's uncertainty propagation.
For the baseline design analysis, we adopt $\chi_w \approx 0.3$--$0.5$ (consistent with the observation that the gas flow removes a substantial fraction of reaction water), yielding $\sigma \approx 1.28$--$1.39$ and $\varepsilon(\xi{=}1) \approx 0.17$--$0.23$. This represents a \textbf{8--25$\times$ increase in flow resistance} over the scrubber's lifetime---significantly more severe than the solid-only estimate and a binding constraint on scrubber sizing and fan power. Both the soda lime and silica gel packed beds contribute to total resistance.
\subsection{Closed-Loop Pressure and Volume Dynamics}
With the counter-lung, the system operates in an approximately \emph{isobaric} regime rather than an isochoric one. The total loop volume is the sum of the rigid components (suit shell, canisters, tubing) and the compliant counter-lung:
\begin{equation}
V_{\mathrm{loop}}(t) = V_{\mathrm{rigid}} + V_{\mathrm{CL}}(t)
\end{equation}
The counter-lung volume evolves according to \cref{eq:counterlung_dynamics}. The fans recirculate gas internally through the treatment train; net mass changes (O\textsubscript{2} injection, CO\textsubscript{2} scrubbing to solid, H\textsubscript{2}O adsorption) are accommodated by counter-lung displacement rather than pressure variation. Small residual pressure oscillations from tidal breathing ($\sim \pm$\SI{1}{\milli\bar}) are smoothed by the counter-lung's compliance.
\subsubsection{Ventilation Sanity Check: Loop Flush Rate vs.\ CO\textsubscript{2} Production}
A basic rebreather design verification ensures that the fan circulation rate is sufficient to keep inspired CO\textsubscript{2} below the \SI{5000}{\ppm} threshold. At heavy exertion, CO\textsubscript{2} production reaches $\dot{n}_{\mathrm{CO_2}} \approx \SI{0.068}{\mole\per\minute}$. The scrubber removes CO\textsubscript{2} at a rate proportional to the inlet concentration and contact time. For the loop to maintain steady-state $p_{\mathrm{CO_2}} \le \SI{5000}{\ppm} = 0.005$~atm, the minimum circulation flow rate must satisfy:
\begin{equation}
Q_{\mathrm{circ,min}} \ge \frac{\dot{n}_{\mathrm{CO_2}} \cdot R_g T}{p_{\mathrm{CO_2,max}} - p_{\mathrm{CO_2,out}}}
\end{equation}
where $p_{\mathrm{CO_2,out}}$ is the scrubber outlet concentration
(effectively zero for fresh soda lime). Evaluating in 
litre-atmosphere units with 
$R = \SI{0.08206}{\liter\,\atm\per\mole\per\kelvin}$ 
at $T = \SI{308}{\kelvin}$ (\SI{35}{\celsius}) and 
$P = \SI{1}{\atm}$:
\begin{equation}
Q_{\mathrm{circ,min}} 
\ge \frac{\dot{n}_{\mathrm{CO_2}} \cdot R T}
         {p_{\mathrm{CO_2,max}} - p_{\mathrm{CO_2,out}}}
= \frac{0.068 \times 0.08206 \times 308}{0.005 - 0}
\approx \SI{344}{\liter\per\minute}
\end{equation} This is a high flow rate but consistent with the $\sim$\SIrange{200}{400}{\liter\per\minute} circulation rates used in military and diving rebreathers under heavy workloads. Accordingly, the loop flow sensor and fan subsystem should be specified to at least \SI{400}{\liter\per\minute} full scale. The fan system must deliver this flow against the total loop pressure drop (scrubber + desiccant + tubing), which is the fan sizing constraint. 

\medskip
\noindent\textbf{Modeling caveat:}
This estimate assumes a perfectly mixed control volume (CSTR approximation) and therefore represents a conservative upper bound on required circulation. In reality, the scrubber is a reactive packed bed where removal efficiency depends on residence time and the Damköhler number rather than simple dilution. Properly sized beds therefore require substantially lower circulation rates than the CSTR bound suggests.
\section{Integrated Mass Balance Summary}
\label{sec:mass_balance}
Collecting the preceding results, the complete state of the closed-loop system is described by a coupled system of ordinary differential equations:
\begin{align}
\frac{dm_{\mathrm{O_2,tank}}}{dt} &= -\dot{m}_{\mathrm{O_2,inject}}(t) \label{eq:state_o2} \\[4pt]
\frac{dn_{\mathrm{O_2,suit}}}{dt} &= \dot{n}_{\mathrm{O_2,inject}}(t) - \dot{n}_{\mathrm{O_2,consumed}}(t) - x_{\mathrm{O_2}}(t)\,\dot{n}_{\mathrm{vent}}(t) \label{eq:state_o2suit} \\[4pt]
\frac{dn_{\mathrm{CO_2,suit}}}{dt} &= \dot{n}_{\mathrm{CO_2,produced}}(t) - r_{\mathrm{scrub}}(t) - x_{\mathrm{CO_2}}(t)\,\dot{n}_{\mathrm{vent}}(t) \label{eq:state_co2} \\[4pt]
\frac{dn_{\mathrm{H_2O,suit}}}{dt} &= \dot{n}_{\mathrm{H_2O,exhaled}}(t) + \dot{n}_{\mathrm{H_2O,rxn}}(t) - \dot{n}_{\mathrm{H_2O,ads}}(t) - x_{\mathrm{H_2O}}(t)\,\dot{n}_{\mathrm{vent}}(t) \label{eq:state_h2o} \\[4pt]
\frac{dn_{\mathrm{N_2,suit}}}{dt} &= -x_{\mathrm{N_2}}(t)\,\dot{n}_{\mathrm{vent}}(t) \label{eq:state_n2} \\[4pt]
\frac{dV_{\mathrm{CL}}}{dt} &= \frac{R_g T_{\mathrm{suit}}}{P_s}\bigl(\dot{n}_{\mathrm{O_2,inject}} - \dot{n}_{\mathrm{O_2,consumed}} - \dot{n}_{\mathrm{vent}}\bigr) + \frac{V_{\mathrm{gas}}}{T_{\mathrm{suit}}}\frac{dT_{\mathrm{suit}}}{dt} + \dot{V}_{\mathrm{breath}}(t) \label{eq:state_cl} \\[4pt]
\frac{dM_{\mathrm{water}}}{dt} &= \dot{m}_{\mathrm{air}} \bigl[ Y_{\mathrm{in}}(t) - Y_{\mathrm{out}}(t) \bigr] \label{eq:state_water} \\[4pt]
\frac{dm_{\mathrm{Ca(OH)_2}}}{dt} &= -r_{\mathrm{scrub}}(t)\, M_{\mathrm{Ca(OH)_2}} \label{eq:state_sorbent} \\[4pt]
\frac{dT_{\mathrm{bed}}}{dt} &= \frac{1}{(\rho c_p)_{\mathrm{bed}} V_{\mathrm{bed}}}
\bigl[
\dot{Q}_{\mathrm{scrub}}(t)
- \dot{m}_{\mathrm{air}} c_{p,\mathrm{air}} (T_{\mathrm{bed}} - T_{\mathrm{air,in}})
- U A_{\mathrm{wall}} (T_{\mathrm{bed}} - T_{\mathrm{wall}})
\bigr]
\label{eq:state_temp} \\[4pt]
\frac{d\mathrm{UPTD}}{dt} &=
\begin{cases}
\displaystyle\frac{1}{60}\,\left(
  \frac{P_{iO_2}(t) - \SI{0.50}{\atm}}{\SI{0.50}{\atm}}
\right)^{0.83},
& \text{if } P_{iO_2}(t) > \SI{0.50}{\atm}, \\[6pt]
0,
& \text{otherwise,}
\end{cases}
\label{eq:state_dose}
\end{align}
where \(t\) is measured in seconds, so the factor \(1/60\) converts the
standard minute-based UPTD accumulation law into SI time units.
The gas-phase water balance (\cref{eq:state_h2o}) couples the humidity state to the silica gel adsorption model: $\dot{n}_{\mathrm{H_2O,exhaled}}$ is the metabolic water vapor production, $\dot{n}_{\mathrm{H_2O,rxn}} = r_{\mathrm{scrub}}(t)$ (1:1 stoichiometry, \cref{eq:scrub_primary}), $\dot{n}_{\mathrm{H_2O,ads}}$ is the molar adsorption rate in the silica gel (\cref{eq:ldf}), and the last term is vent loss proportional to the water vapor mole fraction. The nitrogen balance (\cref{eq:state_n2}) is the simplest ODE---N\textsubscript{2} is neither produced nor consumed, only lost through venting---but it is essential: the total gas-phase molar inventory $n_{\mathrm{total}}(t) = n_{\mathrm{O_2}} + n_{\mathrm{CO_2}} + n_{\mathrm{H_2O}} + n_{\mathrm{N_2}}$ appears in the denominator of every mole-fraction computation, and without tracking $n_{\mathrm{N_2}}$ it cannot be evaluated. The sorbent depletion (\cref{eq:state_sorbent}) tracks calcium hydroxide mass directly, where $r_{\mathrm{scrub}}$ is the molar scrubbing rate (mol~CO\textsubscript{2}/s) and $M_{\mathrm{Ca(OH)_2}} = \SI{74.09}{\gram\per\mole}$; the 1:1 stoichiometry of the net reaction (\cref{eq:scrub_primary}) means each mole of CO\textsubscript{2} scrubbed consumes one mole of Ca(OH)\textsubscript{2}.
The cumulative oxygen toxicity dose $\mathrm{UPTD}$ (\cref{eq:state_dose}) accumulates whenever $P_{iO_2}$ exceeds the pulmonary toxicity threshold of \SI{0.50}{\atm}, following the standard unit pulmonary toxicity dose (UPTD) power-law accumulation model used in diving and hyperbaric medicine. This enables the MPC to trade transient high-O\textsubscript{2} exposure against a cumulative dose budget over the mission, rather than enforcing only an instantaneous threshold.
Subject to the constraints:
\begin{align}
m_{\mathrm{O_2,tank}}(t) &\ge 0 \label{eq:con1}\\
m_{\mathrm{Ca(OH)_2}}(t) &\ge 0 \label{eq:con2}\\
M_{\mathrm{water}}(t) &\le M_{\mathrm{water,max}} \label{eq:con3}\\
\SI{0.19}{\atm} \le P_{iO_2}(t) &\le \SI{0.50}{\atm} \label{eq:con4}\\
x_{\mathrm{O_2}}(t) &\le 0.235 \label{eq:con5}\\
x_{\mathrm{CO_2,suit}}(t) &\le \SI{5000}{\ppm} \label{eq:con6}\\
\mathrm{RH}_{\mathrm{suit}}(t) &\le \SI{60}{\percent} \label{eq:con7}\\
V_{\mathrm{CL}}(t) &\ge V_{\mathrm{CL,min}} \label{eq:con8}\\
\mathrm{UPTD}(t) &\le \mathrm{UPTD}_{\max} \label{eq:con9}
\end{align}
where inspired O\textsubscript{2} partial pressure is computed on a \emph{wet-gas} basis to reflect the physiologically relevant alveolar gas equation:
\begin{equation}
\label{eq:pio2_wet}
\begin{aligned}
P_{iO_2}(t)
&= \Bigl(P_s(t) - P_{\mathrm{H_2O}}\!\bigl(T_{\mathrm{suit,bz}},\,\mathrm{RH}_{\mathrm{suit}}\bigr)\Bigr) \\
&\quad \times x_{\mathrm{O_2}}(t)
\end{aligned}
\end{equation}
The magnitude of this correction depends on conditions at the
breathing zone. At the suit breathing-zone temperature
$T_{\mathrm{suit,bz}} \approx \SI{35}{\celsius}$, the saturation
vapor pressure is $P_{\mathrm{H_2O}}^{\mathrm{sat}} \approx
\SI{42}{\mmHg}$; at the control target of $\mathrm{RH} = 60\%$,
$P_{\mathrm{H_2O}} \approx \SI{25}{\mmHg} \approx
\SI{0.033}{\atm}$, reducing $P_{iO_2}$ by $\sim$3\% relative to a
dry-gas computation. For comparison, the classical alveolar gas
equation uses $P_{\mathrm{H_2O}} = \SI{47}{\mmHg}$ (body core
temperature, \SI{37}{\celsius}, full saturation in the alveolar
space)---the relevant correction when estimating alveolar
$P_A\mathrm{O}_2$ from inspired $P_{iO_2}$, but not applicable to the
inspired-gas calculation here. At the humidity constraint boundary
($\mathrm{RH} = 60\%$), the inspired-gas correction is modest; it
becomes clinically significant if the desiccant saturates and RH
approaches 100\%, where $P_{\mathrm{H_2O}} \to \SI{42}{\mmHg}$
and the correction approaches $\sim$5.5\%. The cumulative toxicity dose constraint (\cref{eq:con9}) limits total pulmonary exposure via the UPTD (unit pulmonary toxicity dose) model, enabling the MPC to trade brief excursions above \SI{0.50}{\atm} against a mission-duration dose budget rather than relying solely on the instantaneous threshold.
$V_{\mathrm{CL,min}}$ is the minimum counter-lung volume required to accommodate tidal breathing oscillations (typically \SIrange{1}{2}{\liter}). The oxygen mole fraction is explicitly constrained by the normal-mode fire-safety limit
$x_{\mathrm{O_2}}(t) \le 0.235$ (\cref{eq:con5}), while the higher toxicity ceiling
of \SI{0.50}{\atm} in wet-basis inspired partial pressure serves as a degraded-mode
backup constraint rather than the primary operating target. The dominant driver of O\textsubscript{2} enrichment is vent compensation (\cref{eq:o2_enrichment_simple}): each vent--refill cycle replaces mixed gas with pure O\textsubscript{2}. The MPC must jointly manage the suit pressure (minimizing unnecessary venting) and O\textsubscript{2} fraction trajectory over its prediction horizon.
\newpage
\part{AI-Based Control System}
\section{Motivation: Why AI Control?}
\label{sec:ai_motivation}
A conventional fixed-setpoint PID (proportional-integral-derivative) controller (as commonly used in rebreather systems) computes its output from three terms---a proportional response to the current error, an integral term that eliminates steady-state offset, and a derivative term that damps overshoot---and regulates each variable independently: O\textsubscript{2} partial pressure, CO\textsubscript{2} concentration, humidity, and fan speed. While adequate for steady-state diving, this approach is fundamentally limited for firefighting because:
\begin{enumerate}[label=(\alph*)]
    \item \textbf{Unknown mission duration:} The firefighter does not know in advance how long they will be deployed. A PID controller cannot anticipate resource depletion and will maintain nominal setpoints until a consumable is exhausted, leading to abrupt system failure.
    \item \textbf{Coupled dynamics:} O\textsubscript{2} injection, CO\textsubscript{2} scrubbing, and humidity are thermodynamically coupled (e.g., faster scrubbing generates more heat and water). Furthermore, vent compensation with pure O\textsubscript{2} creates a direct tradeoff between suit pressure maintenance and oxygen enrichment (\cref{sec:o2_enrichment}) that PID loops operating independently cannot resolve.
    \item \textbf{Rapidly changing conditions:} Fireground conditions---temperature, toxic gas exposure, structural collapse risk---change on timescales of seconds to minutes. The controller must incorporate situational awareness to modify its strategy.
    \item \textbf{Firefighter state variability:} Exertion, stress, and environmental heat load produce highly nonlinear and time-varying metabolic demand that fixed-gain controllers cannot track efficiently.
\end{enumerate}
We adopt an AI control architecture that addresses these challenges through sensor fusion, online state estimation, and model-predictive optimization.
\section{Sensor Suite}
\label{sec:sensors}
The Galactic Bioware Life Support System integrates three categories of sensors: external environmental sensors that characterize the fireground hazard, internal suit environment sensors that monitor the breathing gas and microclimate the firefighter actually experiences, and firefighter biometric sensors that track physiological state. A key design constraint is that all biometric sensors must be compatible with heavy gloves, a fully encapsulating suit, and violent physical activity---ruling out clinical-grade devices such as ingestible core temperature pills, finger/ear pulse oximeters, chest impedance belts, and transcutaneous CO\textsubscript{2} monitors, none of which survive the mechanical and thermal stresses of structural firefighting.
\subsection{External Environmental Sensors}
These sensors face outward through the suit shell or are mounted on the exterior helmet surface, characterizing the ambient hazard environment.
\begin{table}[H]
\centering
\small
\setlength{\tabcolsep}{4pt}
\renewcommand{\arraystretch}{1.08}
\begin{tabularx}{\linewidth}{@{}
>{\raggedright\arraybackslash}X
>{\raggedright\arraybackslash}X
>{\raggedright\arraybackslash}p{0.27\linewidth}
@{}}
\toprule
\textbf{Sensor} & \textbf{Measurement} & \textbf{Range / Resolution} \\
\midrule
Radiant heat flux sensor
& External thermal threat $\dot{q}_{\mathrm{rad}}$
& 0--\SI{200}{\kilo\watt\per\meter\squared} \\

External thermistor (shielded)
& Ambient temperature $T_{\mathrm{ext}}$
& $-$40 to \SI{+500}{\celsius}, $\pm$\SI{2}{\celsius} \\

Toxic gas sensor (MOS array)
& CO, HCN, NO\textsubscript{x} (external)
& Multi-gas, semi-quantitative \\

Barometric pressure sensor
& Ambient pressure $P_a$
& 800--\SI{1100}{\hecto\pascal}, $\pm$\SI{0.5}{\hecto\pascal} \\
\bottomrule
\end{tabularx}
\end{table}

\subsection{Suit Environment Sensors}
These sensors are positioned within the closed breathing loop and the suit interior, providing direct measurement of the gas the firefighter is breathing and the microclimate surrounding their body. Because they operate inside the sealed, positive-pressure envelope, they are shielded from external combustion products and extreme temperatures.

\begin{table}[H]
\centering
\small
\setlength{\tabcolsep}{4pt}
\renewcommand{\arraystretch}{1.08}
\begin{tabularx}{\linewidth}{@{}
>{\raggedright\arraybackslash}X
>{\raggedright\arraybackslash}X
>{\raggedright\arraybackslash}p{0.27\linewidth}
@{}}
\toprule
\textbf{Sensor} & \textbf{Measurement} & \textbf{Range / Resolution} \\
\midrule
NDIR CO\textsubscript{2} sensor
& In-suit $x_{\mathrm{CO_2}}$
& 0--\SI{10}{\percent}, $\pm$\SI{0.01}{\percent} \\

Galvanic O\textsubscript{2} ($\times 3$, median voting)
& In-suit $x_{\mathrm{O_2}}$
& 0--\SI{100}{\percent}, $\pm$\SI{0.1}{\percent} \\

Capacitive RH sensor
& In-suit relative humidity
& 0--\SI{100}{\percent} RH, $\pm$\SI{1.5}{\percent} \\

In-suit thermistor (breathing zone)
& Breathing gas temperature $T_{\mathrm{suit,bz}}$
& 10 to \SI{+60}{\celsius}, $\pm$\SI{0.5}{\celsius} \\

In-suit thermistor (torso)
& Suit interior temperature $T_{\mathrm{suit,torso}}$
& 10 to \SI{+70}{\celsius}, $\pm$\SI{0.5}{\celsius} \\

Thermocouple (K-type)
& Scrubber bed temperature $T_{\mathrm{bed}}$
& $-$40 to \SI{+1000}{\celsius}, $\pm$\SI{1}{\celsius} \\

Differential pressure transducer
& Suit gauge pressure $\Delta P_{\mathrm{suit}}$
& 0--\SI{50}{\milli\bar}, $\pm$\SI{0.1}{\milli\bar} \\

Flow sensor (hot-wire anemometer)
& Loop circulation rate $Q_{\mathrm{circ}}$
& 0--\SI{400}{\liter\per\minute}, $\pm$\SI{2}{\percent} \\

Counter-lung position sensor
& Counter-lung volume $V_{\mathrm{CL}}$
& 0--\SI{10}{\liter}, $\pm$\SI{0.05}{\liter} \\
\bottomrule
\end{tabularx}
\end{table}

The in-suit CO\textsubscript{2} and O\textsubscript{2} sensors are particularly
critical: they provide a direct, real-time measure of the breathing-gas
composition the firefighter is actually inhaling. The three galvanic O\textsubscript{2} cells use median voting with plausibility checks (rejecting readings that drift more than \SI{2}{\percent} from the median), providing fault tolerance against the sensor drift and failure modes that are a dominant accident driver in rebreather systems. The counter-lung position sensor provides information about the system's net molar balance and suit integrity. The dual-location temperature sensors (breathing zone and torso) allow the controller to distinguish between a rise in breathing gas temperature (indicating scrubber exotherm or external heat soak) and a rise in body-proximate temperature (indicating metabolic heat accumulation), informing different control responses.
\subsection{Firefighter Biometric Sensors}
All biometric sensors are designed for compatibility with structural firefighting: ruggedized, sweat-resistant, tolerant of high-g impacts, and requiring no exposed skin contact on fingers, ears, or mucous membranes. The chest-strap ECG and torso-mounted IMU are integrated into the suit's inner garment layer. A ruggedized wrist module worn under the suit glove provides redundant heart rate measurement via photoplethysmography (PPG) and additional motion sensing.
\begin{table}[H]
\centering
\small
\setlength{\tabcolsep}{4pt}
\renewcommand{\arraystretch}{1.08}
\begin{tabularx}{\linewidth}{@{}
>{\raggedright\arraybackslash}X
>{\raggedright\arraybackslash}X
>{\raggedright\arraybackslash}p{0.27\linewidth}
@{}}
\toprule
\textbf{Sensor} & \textbf{Measurement} & \textbf{Range / Resolution} \\
\midrule
Chest-strap ECG (dry electrode)
& Heart rate (HR)
& 30--\SI{240}{\bpm}, $\pm$\SI{1}{\bpm} \\

Chest-strap ECG
& Heart rate variability (HRV)
& R--R intervals, \SI{1}{\milli\second} \\

Torso IMU (accelerometer + gyro)
& Activity level / posture
& 3-axis, \SI{100}{\hertz} \\

Wrist PPG sensor
& Redundant HR, perfusion index
& 30--\SI{240}{\bpm}, $\pm$\SI{3}{\bpm} \\

Wrist accelerometer
& Wrist motion / activity
& 3-axis, \SI{50}{\hertz} \\
\bottomrule
\end{tabularx}
\end{table}
\begin{remark}[On the absence of SpO\textsubscript{2} and $P_{\mathrm{tc}}\mathrm{CO}_2$ sensing]
Clinical pulse oximetry and transcutaneous CO\textsubscript{2} monitoring require stable skin contact, controlled temperature, and minimal motion artifact---conditions fundamentally incompatible with structural firefighting. Without direct blood gas measurement, the Galactic Bioware Life Support System does not attempt to estimate arterial blood gas values (which are poorly identifiable from inspired gas composition alone due to confounders including V/Q mismatch, shunt fraction, and hemoglobin variability). Instead, the system computes \emph{risk indices}:
\begin{itemize}[nosep]
    \item \textbf{Hypoxia risk index}, driven by wet-basis $P_{iO_2}$ (\cref{eq:pio2_wet}), estimated minute ventilation (from HR and activity data), and the metabolic model's O\textsubscript{2} consumption estimate.
    \item \textbf{Hypercapnia risk index}, driven by inspired
    \(x_{\mathrm{CO_2}}\), estimated CO\textsubscript{2} production rate,
    and the scrubber's modeled removal capacity.
\end{itemize}
These risk indices are sufficient for control decisions (the MPC penalizes high risk, not specific blood gas values) and avoid claiming identifiability that the sensor suite cannot support.
\end{remark}
\begin{remark}[On core temperature estimation]
Core body temperature $T_c$ is a critical physiological variable for heat stress management, but direct measurement via ingestible telemetry pills is impractical for routine firefighting deployment (pre-ingestion timing, single-use cost, gastrointestinal concerns). The Galactic Bioware Life Support System instead estimates $T_c$ using a \emph{Kalman-filter--based thermal model} that combines HR, HRV, in-suit torso temperature, activity intensity from IMU, and the known external thermal environment. This approach, validated in military heat stress research \cite{buller_core_temp}, provides $T_c$ estimates within $\pm$\SI{0.3}{\celsius} of pill-measured values under moderate to heavy exertion. Accuracy may degrade to $\pm$\SI{0.5}{\celsius}--\SI{0.7}{\celsius} under extreme heat---precisely the conditions encountered in structural firefighting---which the EKF's uncertainty quantification captures as increased state covariance.
\end{remark}
\section{State-Space Formulation}
\label{sec:state_space}
\subsection{State Vector}
We define the system state vector $\mathbf{x}(t) \in \mathbb{R}^{18}$:
\begin{equation}
\label{eq:state_vec}
\mathbf{x}(t) = \begin{bmatrix}
n_{\mathrm{O_2,suit}} \\[2pt]
n_{\mathrm{CO_2,suit}} \\[2pt]
n_{\mathrm{H_2O,suit}} \\[2pt]
n_{\mathrm{N_2,suit}} \\[2pt]
x_{\mathrm{O_2}} \\[2pt]
V_{\mathrm{CL}} \\[2pt]
m_{\mathrm{O_2,tank}} \\[2pt]
m_{\mathrm{Ca(OH)_2}} \\[2pt]
\xi \\[2pt]
M_{\mathrm{water}} \\[2pt]
T_{\mathrm{bed}} \\[2pt]
T_{\mathrm{suit,bz}} \\[2pt]
T_{\mathrm{suit,torso}} \\[2pt]
\mathrm{HR} \\[2pt]
\hat{T}_c \\[2pt]
\hat{\dot{V}}_{\mathrm{O_2}} \\[2pt]
\hat{W} \\[2pt]
\mathrm{UPTD}
\end{bmatrix}
\end{equation}
The first thirteen states are physical/chemical plant and suit environment states from Part~I. $n_{\mathrm{H_2O,suit}}$ is the gas-phase water inventory in the breathing loop (mol), governed by the balance of exhaled moisture, reaction-generated moisture, silica gel adsorption, and vent losses (\cref{eq:state_h2o}). $n_{\mathrm{N_2,suit}}$ is the nitrogen inventory (mol), which starts at $\sim$79\% of the initial gas fill and is continuously depleted by exhaust-valve venting but never replenished (\cref{eq:state_n2}); tracking $n_{\mathrm{N_2}}$ is essential because the total molar inventory $n_{\mathrm{total}}(t) = n_{\mathrm{O_2}} + n_{\mathrm{CO_2}} + n_{\mathrm{H_2O}} + n_{\mathrm{N_2}}$ cannot be computed without it, and $n_{\mathrm{total}}$ appears in the denominators of all mole-fraction computations. The O\textsubscript{2} mole fraction is constrained by the species inventories:
\begin{equation}
\label{eq:xo2_consistency}
x_{\mathrm{O_2}}(t) =
\frac{n_{\mathrm{O_2,suit}}(t)}
{n_{\mathrm{O_2,suit}}(t)+n_{\mathrm{CO_2,suit}}(t)+n_{\mathrm{H_2O,suit}}(t)+n_{\mathrm{N_2,suit}}(t)}
\end{equation}
In implementation, $x_{\mathrm{O_2}}$ may still be carried in the EKF as a redundant sensed state, but it is tied to the molar inventories by this algebraic consistency relation (or an equivalent pseudo-measurement). $V_{\mathrm{CL}}$ is the counter-lung volume (\cref{eq:counterlung_dynamics}), directly measured by a position sensor; suit pressure $P_s$ is computed algebraically from $V_{\mathrm{CL}}$ via the counter-lung compliance relation (\cref{eq:ps_algebraic}) rather than tracked as an independent state, avoiding a differential-algebraic inconsistency. $m_{\mathrm{Ca(OH)_2}}$ is the remaining mass of calcium hydroxide (the limiting reagent in soda lime), decremented by the 1:1 stoichiometry of the net scrubbing reaction (\cref{eq:state_sorbent}). $\xi$ is the scrubber conversion fraction, which determines the time-varying void fraction $\varepsilon(t)$ via \cref{eq:epsilon_t} and hence the scrubber flow resistance; it is not directly measured but inferred by the EKF from the cumulative scrubbing integral and the observed fan-speed--to--flow-rate relationship. States $T_{\mathrm{suit,bz}}$ and $T_{\mathrm{suit,torso}}$ are the breathing-zone and torso-interior temperatures. The last five are firefighter physiological and safety states: heart rate (directly measured), \emph{estimated} core temperature, estimated oxygen consumption rate, estimated metabolic work rate, and cumulative oxygen toxicity dose $\mathrm{UPTD}$ (\cref{eq:state_dose}).
\subsection{Control Input Vector}
The control input $\mathbf{u}(t) \in \mathbb{R}^3$:
\begin{equation}
\label{eq:control_vec}
\mathbf{u}(t) = \begin{bmatrix}
\dot{m}_{\mathrm{O_2,inject}} \\[2pt]
\omega_{\mathrm{fan}} \\[2pt]
\phi_{\mathrm{bypass}}
\end{bmatrix}
\end{equation}
where $\dot{m}_{\mathrm{O_2,inject}}$ is the oxygen injection mass flow rate (proportional valve opening), $\omega_{\mathrm{fan}}$ is the fan rotational speed (controlling circulation rate), and $\phi_{\mathrm{bypass}} \in [0,1]$ is the fraction of flow bypassing the scrubber (a controllable damper), which allows trading scrubbing rate against pressure drop and thermal load.
\subsubsection{Solenoid Valve Stiction and Low-Flow Nonlinearity}
The proportional solenoid valve that meters O\textsubscript{2} injection exhibits a nonlinear dead-zone at low command signals due to \emph{stiction} (static friction) of the valve armature against its seat. Below a threshold voltage $V_{\mathrm{break}}$, the valve does not move; once the breakaway force is exceeded, the valve jumps to a minimum open position $\dot{m}_{\min}$, creating a discontinuity in the control-to-flow mapping:
\begin{equation}
\label{eq:stiction}
\dot{m}_{\mathrm{O_2,actual}}(V) = \begin{cases}
0 & \text{if } V < V_{\mathrm{break}} \\
\dot{m}_{\min} + k_v(V - V_{\mathrm{break}}) & \text{if } V \ge V_{\mathrm{break}}
\end{cases}
\end{equation}
where $V$ is the command voltage and $k_v$ is the valve gain in the linear regime. The jump from 0 to $\dot{m}_{\min}$ (typically \SIrange{5}{15}{\percent} of full-scale flow) is non-differentiable, which creates two problems for the controller:
\begin{enumerate}[nosep]
    \item The MPC's gradient-based solver (SQP) cannot compute a valid descent direction at the stiction boundary, leading to oscillation between ``valve closed'' and ``valve at minimum open.''
    \item At very low metabolic demand (rest periods), the desired injection rate may fall below $\dot{m}_{\min}$ regardless of the resource cost formulation. The controller ``hunts'' for the unreachable setpoint, causing high-frequency chatter against the valve seat---accelerating exactly the mechanical fatigue that the RL cycling penalty was designed to prevent.
\end{enumerate}

\medskip
\noindent\textbf{Pulse-width modulation (PWM) strategy for sub-minimum flows:}
The standard solution in precision fluid control is to replace continuous low-voltage commands with a \emph{pulse-width modulated} (PWM) signal: the valve is periodically opened to a flow rate $\dot{m}_{\mathrm{pulse}} > \dot{m}_{\min}$ (safely above the stiction boundary) for a fraction $\delta(t) \in [0,1]$ of each PWM period $T_{\mathrm{PWM}}$, achieving the desired time-averaged flow:
\begin{equation}
\label{eq:pwm}
\langle \dot{m}_{\mathrm{O_2,inject}} \rangle = \delta(t) \cdot \dot{m}_{\mathrm{pulse}}
\end{equation}
The MPC optimizes the \emph{duty cycle} $\delta(t)$ rather than the raw voltage when the desired flow falls below $\dot{m}_{\min}$. This keeps the valve operating in its linear regime (above $V_{\mathrm{break}}$) during each pulse, eliminating the stiction nonlinearity from the optimization landscape. The PWM period $T_{\mathrm{PWM}} \approx \SIrange{2}{5}{\second}$ is chosen to be:
\begin{itemize}[nosep]
    \item Long enough that each open--close cycle keeps the valve in steady 
    flow (avoiding water-hammer transients),
    \item Short enough that the per-pulse O\textsubscript{2} injection is a 
    small fraction of the loop inventory, bounding the mole-fraction 
    fluctuation (see below),
    \item Below the \SI{2}{\hertz} reversal threshold in the RL cycling 
    penalty, so PWM operation is not penalized as ``chatter.''
\end{itemize}

\medskip
\noindent\textbf{Bounding the PWM-induced O\textsubscript{2} fluctuation:}
Each pulse injects a bolus of $\Delta n_{\mathrm{pulse}} = 
\dot{m}_{\mathrm{pulse}} \cdot \delta \cdot T_{\mathrm{PWM}} / 
M_{\mathrm{O_2}}$ moles of pure O\textsubscript{2} into a loop containing 
$n_{\mathrm{total}} \approx \SI{4}{\mole}$ ($\sim$\SI{100}{\liter} at 
\SI{1}{\atm}, \SI{308}{\kelvin}). At worst case---full-scale pulse flow 
$\dot{m}_{\mathrm{pulse}} = \SI{1}{\gram\per\second}$, duty cycle 
$\delta = 1$, and $T_{\mathrm{PWM}} = \SI{5}{\second}$---the injected 
bolus is:
\begin{equation}
\Delta n_{\mathrm{pulse}} = \frac{1.0 \times 5}{32.0} 
\approx \SI{0.156}{\mole}
\end{equation}
The resulting peak-to-trough mole-fraction excursion, assuming 
instantaneous injection into a well-mixed volume (worst case; in practice 
the fan circulation provides continuous mixing during the pulse), is:
\begin{equation}
\Delta x_{\mathrm{O_2}} \approx 
\frac{\Delta n_{\mathrm{pulse}}}{n_{\mathrm{total}} 
+ \Delta n_{\mathrm{pulse}}} (1 - x_{\mathrm{O_2}})
= \frac{0.156}{4.156}(1 - 0.21) \approx 0.030 
= \SI{3.0}{\percent}
\end{equation}
This is the \emph{unmixed bolus} upper bound. In practice, the fan 
circulation rate ($\sim$\SI{300}{\liter\per\minute}$\gg$ loop volume / 
$T_{\mathrm{PWM}}$) turns over the loop volume multiple times during each 
\SI{5}{\second} pulse, distributing the injected O\textsubscript{2} 
throughout the loop as it enters. The relevant mixing timescale is 
$\tau_{\mathrm{mix}} = V_{\mathrm{loop}}/Q_{\mathrm{circ}} \approx 
100/300 \approx \SI{0.33}{\minute} \approx \SI{20}{\second}$, and the 
pulse duration is a fraction of this, so the instantaneous local 
fluctuation at the breathing zone is attenuated by approximately 
$T_{\mathrm{PWM}} \cdot \delta / (2\tau_{\mathrm{mix}})$. For a 
typical low-demand duty cycle $\delta \approx 0.3$ and 
$T_{\mathrm{PWM}} = \SI{3}{\second}$:
\begin{equation}
\Delta x_{\mathrm{O_2,bz}} \approx 
\frac{\dot{m}_{\mathrm{pulse}} \cdot \delta \cdot T_{\mathrm{PWM}}}
{M_{\mathrm{O_2}} \cdot n_{\mathrm{total}}} (1 - x_{\mathrm{O_2}})
= \frac{1.0 \times 0.3 \times 3}{32.0 \times 4}(0.79)
\approx 0.006 = \SI{0.6}{\percent}
\end{equation}
The $< \SI{0.5}{\percent}$ target is therefore achievable for duty cycles 
$\delta \lesssim 0.25$ at $T_{\mathrm{PWM}} = \SI{3}{\second}$, which 
corresponds to the low-demand regime where PWM is actually used (higher 
demands use continuous flow above the stiction threshold). For the 
separated-loop architecture ($V_{\mathrm{loop}} \approx 
\SI{10}{\liter}$, $n_{\mathrm{total}} \approx \SI{0.4}{\mole}$), the 
fluctuation is proportionally larger and $T_{\mathrm{PWM}}$ must be 
reduced accordingly.
The transition between continuous control (above $\dot{m}_{\min}$) and PWM control (below $\dot{m}_{\min}$) is managed by a hysteresis band to avoid mode-switching oscillation at the boundary.
\subsection{Disturbance Vector}
The uncontrolled disturbances $\mathbf{d}(t)$ include:
\begin{equation}
\mathbf{d}(t) = \begin{bmatrix}
W(t) & T_{\mathrm{ext}}(t) & \dot{q}_{\mathrm{rad}}(t) & c_{\mathrm{toxic}}(t) & P_a(t)
\end{bmatrix}^T
\end{equation}
representing metabolic work rate, external temperature, radiant heat flux, external toxic gas concentrations, and ambient barometric pressure.
\subsection{Nonlinear State Dynamics}
The system dynamics are:
\begin{equation}
\label{eq:dynamics}
\dot{\mathbf{x}}(t) = \mathbf{f}\bigl(\mathbf{x}(t), \mathbf{u}(t), \mathbf{d}(t)\bigr)
\end{equation}
where $\mathbf{f}$ encapsulates the coupled ODEs from \crefrange{eq:state_o2}{eq:state_temp} plus suit environment dynamics and physiological models:

\medskip
\noindent\textbf{Suit breathing-zone temperature dynamics:}
\begin{equation}
C_{\mathrm{bz}} \frac{dT_{\mathrm{suit,bz}}}{dt} = \dot{m}_{\mathrm{air}} c_{p,\mathrm{air}} (T_{\mathrm{bed,out}} - T_{\mathrm{suit,bz}}) - h_{\mathrm{bz}} A_{\mathrm{bz}} (T_{\mathrm{suit,bz}} - T_{\mathrm{suit,torso}})
\end{equation}
where $C_{\mathrm{bz}}$ is the thermal capacitance of the breathing zone gas volume and $T_{\mathrm{bed,out}}$ is the scrubber outlet air temperature---capturing the direct effect of scrubber exotherm on inspired gas temperature.

\medskip
\noindent\textbf{Suit torso-interior temperature dynamics:}
\begin{equation}
C_{\mathrm{torso}} \frac{dT_{\mathrm{suit,torso}}}{dt} = \dot{Q}_{\mathrm{met,skin}}(\hat{W}) + U_{\mathrm{shell}} A_{\mathrm{shell}} (T_{\mathrm{ext}} - T_{\mathrm{suit,torso}}) + \dot{q}_{\mathrm{rad}} \tau_{\mathrm{shell}} A_{\mathrm{shell}} - \dot{Q}_{\mathrm{conv,suit}}
\end{equation}
where $\dot{Q}_{\mathrm{met,skin}}$ is metabolic heat reaching the skin surface, $U_{\mathrm{shell}}$ is the suit shell's overall thermal transmittance, $\tau_{\mathrm{shell}}$ is the shell's radiant transmissivity, and $\dot{Q}_{\mathrm{conv,suit}}$ is convective cooling from the circulating gas flow. The rise rate of $T_{\mathrm{suit,torso}}$ is a key observable for the controller: a rapid increase indicates either high external heat soak or rising metabolic heat load.

\medskip
\noindent\textbf{Heart rate dynamics} (first-order lag model):
\begin{equation}
\tau_{\mathrm{HR}} \frac{d\mathrm{HR}}{dt} = \mathrm{HR}_{\mathrm{ss}}(\hat{W}, \hat{T}_c, T_{\mathrm{ext}}) - \mathrm{HR}(t)
\end{equation}
\noindent\textbf{Estimated core temperature dynamics} (Kalman-filter--augmented Stolwijk model \cite{stolwijk}):
\begin{equation}
C_c \frac{d\hat{T}_c}{dt} = \dot{Q}_{\mathrm{met}}(\hat{W}) - \dot{Q}_{\mathrm{resp}} - \dot{Q}_{\mathrm{skin}}(\hat{T}_c, T_{\mathrm{suit,torso}}, \dot{q}_{\mathrm{rad}}) + K_{\mathrm{tc}} \bigl[ T_{\mathrm{suit,torso}}^{\mathrm{meas}} - h(\hat{T}_c) \bigr]
\end{equation}
where the last term is a Kalman correction: $h(\hat{T}_c)$ is the predicted torso-interior temperature given the estimated core temperature, and $K_{\mathrm{tc}}$ is the Kalman gain. This fuses the physics-based thermal model with the measured in-suit torso temperature to continuously update the core temperature estimate without requiring an ingestible pill.

\medskip
\noindent\textbf{Metabolic estimator with thermal and hypoxic decoupling:}
Heart rate elevation in a firefighter has three distinct physiological causes: (i)~muscular work (the metabolic signal the controller needs), (ii)~cardiovascular heat strain from high humidity and temperature (which impairs evaporative cooling and triggers compensatory tachycardia), and (iii)~hypoxic compensation (chemoreceptor-driven HR increase when $P_{iO_2}$ drops below $\sim$\SI{0.18}{\atm}). If the metabolic estimator treats all HR elevation as muscular work, two dangerous positive feedback loops emerge:
\emph{Humidity--scrubbing loop:} High suit humidity $\to$ impaired sweat evaporation $\to$ elevated HR $\to$ model over-estimates $\hat{W}$ $\to$ controller increases O\textsubscript{2} injection and fan speed $\to$ more CO\textsubscript{2} scrubbed per unit time $\to$ more scrubber heat and reaction-generated moisture $\to$ higher humidity. This loop is particularly dangerous when heat reduces the silica gel's adsorption capacity ($q_m$ in \cref{eq:gab} drops approximately 40\% between \SI{25}{\celsius} and \SI{50}{\celsius}).
\emph{Hypoxia--conservation loop:} $\lambda(t)$ forces O\textsubscript{2} 
fraction down $\to$ mild hypoxia elevates HR via peripheral chemoreceptor 
activation \cite{marshall_chemo} $\to$ model over-estimates $\hat{W}$ $\to$ controller increases O\textsubscript{2} to ``support exertion'' $\to$ conflicts with scarcity multiplier $\to$ valve oscillation between conservation and support.
To break both loops, the metabolic estimator receives the in-suit humidity and O\textsubscript{2} fraction as explicit inputs, enabling it to learn the non-work components of HR:
\begin{equation}
\hat{W}(t) = g\bigl(\mathrm{HR}(t), \mathrm{HRV}(t), \mathrm{acc}_{\mathrm{torso}}(t), \mathrm{acc}_{\mathrm{wrist}}(t), T_{\mathrm{suit,torso}}(t), x_{\mathrm{CO_2,suit}}(t), \underbrace{\mathrm{RH}_{\mathrm{suit}}(t), \; x_{\mathrm{O_2}}(t)}_{\text{decoupling inputs}}\bigr)
\end{equation}
The network architecture decomposes the HR signal internally:
\begin{equation}
\label{eq:hr_decomp}
\mathrm{HR}(t) = \underbrace{\mathrm{HR}_{\mathrm{work}}(\hat{W})}_{\text{muscular}} + \underbrace{\mathrm{HR}_{\mathrm{heat}}(\mathrm{RH}, T_{\mathrm{suit,torso}}, \hat{T}_c)}_{\text{heat strain}} + \underbrace{\mathrm{HR}_{\mathrm{hypox}}(x_{\mathrm{O_2}})}_{\text{hypoxic}}
\end{equation}
Only $\mathrm{HR}_{\mathrm{work}}$ drives the metabolic work estimate $\hat{W}$. The heat-strain and hypoxic components are estimated by the network's internal representation and subtracted before the work-rate output layer. During pre-training, the network is exposed to laboratory protocols that independently manipulate humidity (climate chamber), O\textsubscript{2} fraction (altitude simulation), and work rate (treadmill), providing the supervised signal to disentangle these three HR drivers.
The MPC cost function includes an additional \emph{thermal decoupling penalty} that prevents the optimizer from responding to humidity-induced HR elevation with increased scrubbing:
\begin{equation}
\label{eq:thermal_decouple}
\ell_{\mathrm{decouple}}(\mathbf{x}, \mathbf{u}) = w_9 \cdot \max\!\left(0, \; \mathrm{RH}_{\mathrm{suit}} - \mathrm{RH}_{\mathrm{thresh}}\right) \cdot \Delta\omega_{\mathrm{fan}}^+
\end{equation}
where $\Delta\omega_{\mathrm{fan}}^+ = \max(0, \omega_{\mathrm{fan},i} - \omega_{\mathrm{fan},i-1})$ is the positive fan-speed increment. This term penalizes fan-speed \emph{increases} specifically when humidity is already elevated, breaking the positive feedback loop by preventing the controller from ``chasing'' a humidity-driven HR signal with more scrubbing activity.
\section{Control Objective: Constrained Optimization Under Uncertainty}
\label{sec:objective}
The fundamental challenge is to maximize the firefighter's operational effectiveness over a mission of \emph{unknown} duration $T$, subject to hard safety constraints. We formulate this as a receding-horizon optimal control problem.
\subsection{Cost Function}
At each control step $k$, with sampling period $\Delta t$, the controller solves:
\begin{equation}
\label{eq:cost}
\begin{aligned}
\min_{\mathbf{u}_{k:k+N}} \; J
= \sum_{i=k}^{k+N} \Bigl(
& w_1 \,\ell_{\mathrm{safety}}(\mathbf{x}_i)
+ w_2 \,\ell_{\mathrm{comfort}}(\mathbf{x}_i)
+ w_3 \,\ell_{\mathrm{resource}}(\mathbf{x}_i,\mathbf{u}_i) \\
& + w_4 \,\|\Delta \mathbf{u}_i\|^2
+ \ell_{\mathrm{decouple}}(\mathbf{x}_i,\mathbf{u}_i)
\Bigr)
\end{aligned}
\end{equation}
where $N$ is the prediction horizon, $\Delta \mathbf{u}_i = \mathbf{u}_{i} - \mathbf{u}_{i-1}$, and:
\textbf{Safety penalty} (barrier function):
\begin{equation}
\ell_{\mathrm{safety}}(\mathbf{x}) = \sum_{j \in \mathcal{S}} \left[\max\!\left(0, \; \frac{x_j - x_j^{\max}}{x_j^{\max} - x_j^{\mathrm{nom}}}\right)^{\!2} + \max\!\left(0, \; \frac{x_j^{\min} - x_j}{x_j^{\mathrm{nom}} - x_j^{\min}}\right)^{\!2}\,\right]
\end{equation}
enforcing soft constraints on $P_{iO_2}$, $x_{\mathrm{O_2}}$, $x_{\mathrm{CO_2}}$, RH, $V_{\mathrm{CL}}$, $\hat{T}_c$, $T_{\mathrm{suit,torso}}$, HR, and scrubber temperature with quadratic penalty as states approach limits.

\medskip
\noindent\textbf{Comfort cost:} Penalizes deviations from ideal breathing conditions (e.g., RH of \SI{40}{\percent}, $T_{\mathrm{suit}}$ of \SI{28}{\celsius}).

\medskip
\noindent\textbf{Resource conservation:}
\begin{equation}
\ell_{\mathrm{resource}}(\mathbf{x}, \mathbf{u}) = \lambda(t) \cdot \dot{n}_{\mathrm{vent}}(\mathbf{x})
\end{equation}
where $\dot{n}_{\mathrm{vent}}(\mathbf{x})$ is the exhaust-valve vent rate (\cref{eq:vent_rate}), which is a function of the suit pressure state (and hence of $V_{\mathrm{CL}}$ via \cref{eq:ps_algebraic}). The penalty targets \emph{venting} rather than injection because the metabolic O\textsubscript{2} demand is fixed and unavoidable---penalizing injection merely forces the optimizer to delay it, driving the counter-lung to $V_{\mathrm{CL,min}}$ and creating pathological bang-bang valve chatter. The actual controllable waste is vented gas, which carries O\textsubscript{2} (at the current enriched fraction) irreversibly out of the suit. By penalizing the vent rate, the MPC learns to maintain suit pressure just below the cracking threshold, minimizing vent losses while preserving the positive-pressure integrity margin.
$\lambda(t)$ is a \emph{dynamic resource scarcity multiplier} that increases as consumables deplete:
\begin{equation}
\label{eq:lambda}
\lambda(t) = \lambda_0 \left( \frac{m_{\mathrm{O_2,tank}}(0)}{m_{\mathrm{O_2,tank}}(t)} \right)^\alpha
\end{equation}
with $\alpha > 1$ creating increasingly aggressive conservation as the tank empties. This is the key mechanism by which the controller handles unknown mission duration: as resources diminish, the optimizer automatically shifts to a more conservative pressure-management regime, tolerating slightly lower suit gauge pressure to reduce vent frequency.
\subsection{Hard Constraints}
The optimization is subject to:
\begin{align}
\mathbf{x}_{i+1} &= \mathbf{A}_{d,i}\,\mathbf{x}_i + \mathbf{B}_{d,i}\,\mathbf{u}_i + \mathbf{g}_{d,i} & \forall i \in [k, k+N] \label{eq:dyn_con} \\
\mathbf{u}_{\min} &\le \mathbf{u}_i \le \mathbf{u}_{\max} & \forall i \label{eq:act_con} \\
m_{\mathrm{O_2,tank},i} &\ge 0 & \forall i \label{eq:tank_con} \\
m_{\mathrm{Ca(OH)_2},i} &\ge 0 & \forall i \label{eq:naoh_con} \\
V_{\mathrm{CL},i} &\ge V_{\mathrm{CL,min}} & \forall i \quad \text{(breathing margin)} \label{eq:cl_con} \\
x_{\mathrm{O_2},i} &\le x_{\mathrm{O_2,max}}(m_i) & \forall i \label{eq:o2_fire_con} \\
P_{iO_2,i} &\ge \SI{0.16}{\atm} & \forall i \quad \text{(hard hypoxia limit, wet basis)} \label{eq:o2_hard} \\
\mathrm{UPTD}_i &\le \mathrm{UPTD}_{\max} & \forall i \quad \text{(cumulative toxicity dose)} \label{eq:dose_con}
\end{align}

where the mode-dependent oxygen upper bound is
\[
x_{\mathrm{O_2,max}}(m_i)=
\begin{cases}
0.235, & m_i=\text{normal mode},\\
0.50,  & m_i=\text{degraded mode}.
\end{cases}
\]
Under normal operation, the binding upper oxygen constraint is
\(x_{\mathrm{O_2},i} \le 0.235\). In a declared degraded mode, the upper
bound may be temporarily relaxed to \(x_{\mathrm{O_2},i} \le 0.50\), but
this must be treated as an emergency fallback with immediate evacuation
alarm rather than a normal operating regime. The enrichment is driven by
exhaust-valve venting (\cref{eq:o2_enrichment_simple}): each vent event
replaces mixed gas with pure O\textsubscript{2}. The MPC must therefore plan
pressure control to minimize unnecessary venting while maintaining the
positive-pressure margin against toxic infiltration.

\medskip
\noindent\textbf{Discretization of the LTV prediction model:}
The dynamic constraint \eqref{eq:dyn_con} is written in terms of the
LTV matrices $(\mathbf{A}_{d,i}, \mathbf{B}_{d,i}, \mathbf{g}_{d,i})$
obtained by exact zero-order-hold discretization of the Jacobian
linearization at each predicted state:
$\mathbf{A}_{d,i} = e^{A_i \Delta t}$,
$\mathbf{B}_{d,i} = \left(\int_0^{\Delta t} e^{A_i \tau}\,d\tau\right) B_i$,
and $\mathbf{g}_{d,i}$ is the affine residual from the linearization
point. In implementation, $\mathbf{B}_{d,i}$ is evaluated via an
augmented matrix exponential, which remains valid even when $A_i$ is
singular. Because the coupled system exhibits stiffness---the
scrubbing kinetics and gas-mixing time constants
($\tau_{\mathrm{mix}} \approx \SI{20}{\second}$) are fast relative
to the thermal and sorbent-depletion dynamics
($\tau \sim 10^2$--$10^3$~s)---explicit forward Euler
discretization of the full nonlinear model would risk numerical
instability at $\Delta t = \SI{1}{\second}$. The matrix-exponential
discretization is unconditionally stable for any $\Delta t$ and
preserves the eigenvalue structure of the continuous-time
linearization, making it the appropriate choice for the embedded
QP formulation. The EKF's own prediction step uses fourth-order
Runge--Kutta (RK4) on the full nonlinear model $\mathbf{f}$,
which is not subject to the same computational constraint as the
MPC horizon rollout.
\section{AI Architecture}
\label{sec:architecture}

\begin{figure}[H]
    \centering
    \includegraphics[width=0.95\textwidth]{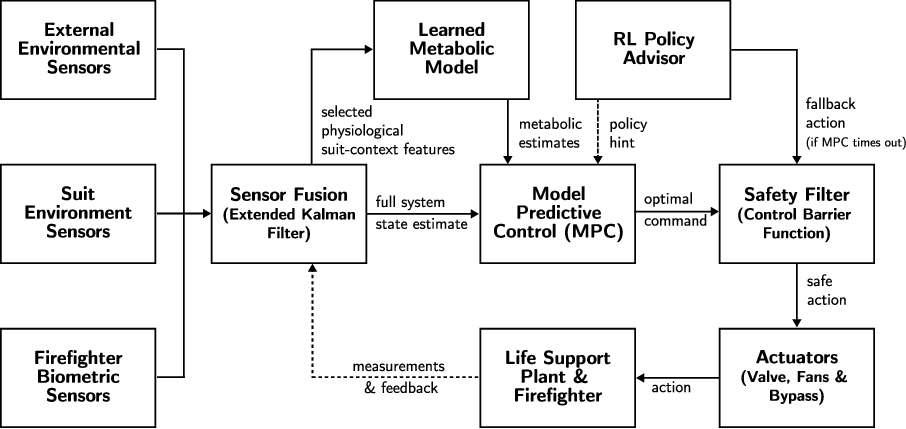}
    \caption{AI control architecture with explicit safety gating. External, in-suit, and biometric sensors feed the EKF-based sensor-fusion and state-estimation layer. The learned metabolic model provides physiological estimates to the MPC, while the RL policy advisor supplies a warm-start policy hint during nominal operation and a fallback candidate action if the MPC fails or times out. All candidate actuator commands pass through the control-barrier-function safety filter before reaching the actuators and, through them, the physical life-support plant and firefighter.}
    \label{fig:architecture}
\end{figure}

The architecture (\cref{fig:architecture}) comprises five key modules, with the safety filter acting as the final supervisory layer between the decision logic and the actuators:
\subsection{Sensor Fusion and State Estimation}
An \emph{extended Kalman filter} (EKF) \cite{simon_ekf} estimates the full state vector $\hat{\mathbf{x}}(t)$ from noisy, asynchronous measurements across all three sensor tiers. The EKF process model uses the nonlinear dynamics $\mathbf{f}$ from \cref{eq:dynamics}, while the measurement model maps states to sensor observations with known noise covariances. Critically, states that are not directly measured---$\hat{T}_c$, $\hat{W}$, and $\hat{\dot{V}}_{\mathrm{O_2}}$---are estimated as latent variables, with their uncertainty explicitly tracked in the state error covariance $\hat{\mathbf{P}}(t)$. The MPC uses this uncertainty quantification for robust constraint satisfaction: when the EKF is less certain about core temperature (e.g., during rapid transients), the controller applies wider safety margins.
\subsection{Learned Metabolic Model}
A compact neural network $g_\theta$, parameterized by $\theta$, maps observable biometric and suit-environment signals to metabolic work rate and O\textsubscript{2} consumption, with explicit \emph{thermal and hypoxic decoupling inputs}:
\begin{equation}
[\hat{W}(t), \, \hat{\dot{V}}_{\mathrm{O_2}}(t)] = g_\theta\bigl(\mathrm{HR}, \mathrm{HRV}, \mathrm{acc}_{\mathrm{torso}}, \mathrm{acc}_{\mathrm{wrist}}, T_{\mathrm{suit,torso}}, x_{\mathrm{CO_2,suit}}, \mathrm{RH}_{\mathrm{suit}}, x_{\mathrm{O_2}}\bigr)
\end{equation}
The network is pre-trained on laboratory data collected from firefighters performing standardized exercises (stair climb, hose advance, forced entry, search and rescue) while instrumented with a metabolic cart, including protocols that independently vary humidity (climate chamber) and inspired O\textsubscript{2} fraction (altitude simulation) to provide supervised signal for the HR decomposition (\cref{eq:hr_decomp}).

\medskip
\noindent\textbf{Online adaptation with catastrophic-forgetting protection:}
The model is fine-tuned online using the EKF's residuals as a self-supervised signal, but na\"ive online gradient descent risks \emph{catastrophic forgetting}: the extreme, non-stationary signal distribution on the fireground (rapid HR spikes, thermal transients, novel exertion patterns) can overwrite the baseline physiological calibrations learned during pre-training. If this occurs, the HR decomposition (\cref{eq:hr_decomp}) degrades---the network may permanently misattribute thermal strain as muscular work, even after the EKF decoupling inputs attempt correction.
To prevent this, online updates use \emph{Elastic Weight Consolidation} (EWC). Let $\theta^*$ denote the pre-trained weights and $\mathbf{F}$ the diagonal of the Fisher information matrix computed on the pre-training dataset, which measures how sensitive the loss is to each weight. The online learning objective augments the self-supervised loss $\mathcal{L}_{\mathrm{EKF}}$ with a quadratic anchor:
\begin{equation}
\label{eq:ewc}
\mathcal{L}_{\mathrm{total}}(\theta) = \mathcal{L}_{\mathrm{EKF}}(\theta) + \frac{\lambda_{\mathrm{EWC}}}{2} \sum_i F_i \,(\theta_i - \theta_i^*)^2
\end{equation}
where $\lambda_{\mathrm{EWC}}$ controls the consolidation strength. Weights with high Fisher information (those critical to the pre-trained HR$\to$work mapping) are strongly anchored to $\theta^*$, while weights with low Fisher information (those encoding context-specific adaptations like individual cardiovascular fitness) are free to update. This preserves the fundamental physiological laws learned in the lab while allowing the model to adapt to the individual firefighter's physiology and to slow sensor drift.
The online learning rate is additionally clamped to $\eta_{\mathrm{online}} \le 0.1 \times \eta_{\mathrm{pretrain}}$, and updates are suspended entirely when the EKF's state uncertainty $\hat{\mathbf{P}}(t)$ exceeds a threshold (indicating that the residual signal is unreliable due to rapid transients).
The inclusion of $\mathrm{RH}_{\mathrm{suit}}$ and $x_{\mathrm{O_2}}$ as inputs is critical: without them, humidity-driven tachycardia and hypoxic compensation are misattributed as muscular exertion, triggering the positive feedback loops described in the state dynamics section. Similarly, in-suit CO\textsubscript{2} concentration serves as a direct metabolic proxy (residual rise above scrubber prediction indicates increased CO\textsubscript{2} production), and the wrist accelerometer provides independent confirmation of upper-body exertion that the torso IMU alone may underestimate.
\subsection{Model-Predictive Controller (MPC)}
At each control interval $\Delta t$ (nominally \SI{1}{\second}), the MPC:
\begin{enumerate}[nosep]
    \item Receives $\hat{\mathbf{x}}_k$ and $\hat{\mathbf{P}}_k$ from the EKF.
    \item Forecasts disturbances $\hat{\mathbf{d}}_{k:k+N}$ using current sensor trends and a short-horizon extrapolation.
    \item Solves the constrained optimization (\cref{eq:cost}--\cref{eq:o2_hard}) over horizon $N$ (typically 15--20 steps with move blocking, i.e., 15--20 seconds ahead) using a linear time-varying (LTV) approximation of the plant dynamics \cite{rawlings_mpc}.
    \item Passes only the first control action $\mathbf{u}_k^*$ through the control-barrier-function safety filter, applies the resulting safe action to the actuators, and re-solves at the next step (receding horizon).
\end{enumerate}
The prediction horizon of 15--20~seconds is short relative to the thermal and sorbent-depletion time constants ($\tau \sim 10^2$--$10^3$~s). The dynamic scarcity multiplier $\lambda(t)$ in \cref{eq:lambda} compensates for this by encoding long-horizon resource awareness into the instantaneous cost structure, effectively extending the controller's planning capability well beyond its optimization horizon. $\lambda(t)$ is the critical innovation: as consumables deplete, the cost of venting increases automatically, causing the optimizer to find operating points that minimize unnecessary gas loss---tightening the pressure margin above cracking pressure (reducing vent frequency and O\textsubscript{2} waste), reducing fan speed where possible (saving O\textsubscript{2} by reducing respiratory demand from increased effort of breathing against loop resistance), and increasing scrubber bypass to extend sorbent life.
\subsection{Reinforcement Learning Policy Advisor}
The MPC optimization is a nonlinear program that may converge slowly or to local minima. To warm-start the optimizer and provide a fallback candidate action for truly novel situations, a reinforcement learning (RL) agent \cite{sutton_rl} runs in parallel. The RL agent is trained offline in a high-fidelity simulator of the Galactic Bioware Life Support System across thousands of randomized fire scenarios with varying:
\begin{itemize}[nosep]
    \item Mission durations (30 min to 4 hours)
    \item Exertion profiles (low, moderate, high, intermittent burst)
    \item Ambient temperature trajectories
    \item Consumable initial states (simulating partial depletion from prior use)
\end{itemize}
The RL policy $\pi_\phi(\mathbf{x})$ provides a ``policy hint'' that can be used to warm-start the MPC during nominal operation. The MPC's nominal command is then passed through the \emph{same safety filter} before reaching the actuators. If the MPC optimization fails or times out (exceeding the \SI{100}{\milli\second} real-time deadline), the RL policy instead supplies a fallback candidate action, which is passed through that same safety filter before application.

\medskip
\noindent\textbf{Control barrier function safety filter:}
Because the RL policy is trained offline and the MPC relies on local linearizations of a nonlinear plant, neither candidate command should be sent directly to the hardware without a final safety check. A \emph{control barrier function} (CBF) filter \cite{ames_cbf} is therefore placed as the last layer before the actuators, ensuring that all hard safety constraints are respected regardless of whether the candidate action comes from the MPC or from the RL fallback path:
\begin{equation}
\label{eq:u_cand}
\mathbf{u}_{\mathrm{cand}} =
\begin{cases}
\mathbf{u}_{\mathrm{MPC}}^* & \text{if the MPC returns a valid solution} \\[4pt]
\pi_\phi(\mathbf{x}) & \text{if the MPC fails or times out}
\end{cases}
\end{equation}

\begin{equation}
\label{eq:cbf}
\mathbf{u}_{\mathrm{safe}} = \arg\min_{\mathbf{u}} \; \|\mathbf{u} - \mathbf{u}_{\mathrm{cand}}\|^2
\quad \text{s.t.} \quad
\dot{h}_j(\mathbf{x}, \mathbf{u}) + \kappa_j \, h_j(\mathbf{x}) \ge 0
\;\; \forall j \in \mathcal{C}
\end{equation}
where $h_j(\mathbf{x})$ are barrier functions for each hard constraint (e.g., $h_1 = x_{\mathrm{O_2,fire}} - x_{\mathrm{O_2}}$, $h_2 = P_{iO_2} - P_{iO_2,\min}$), $\kappa_j > 0$ are class-$\mathcal{K}$ function coefficients, and $\mathcal{C}$ is the set of safety constraints. The CBF filter is a small quadratic program that can be solved in $< \SI{1}{\milli\second}$, ensuring that the system always remains within the safe invariant set even when the RL policy produces an aggressive or out-of-distribution suggestion. The continuous-time CBF condition $\dot{h}_j + \kappa_j h_j \ge 0$ is applied at the QP's sub-millisecond solve cadence; for the embedded implementation at $\Delta t = \SI{1}{\second}$, the equivalent discrete-time condition $h_j(\mathbf{f}(\mathbf{x},\mathbf{u})) \ge (1 - \kappa_j) h_j(\mathbf{x}_k)$ is used to account for the sampling interval.
This architecture ensures that the RL policy is \emph{never} the direct actuator authority: in nominal operation it provides only a warm-start hint to the MPC, and in fallback operation it provides a substitute candidate action when the MPC fails or times out. In both cases, no command reaches the actuators without first passing through the CBF safety layer. The safety filter therefore acts as the final invariant-set guard, providing formal guarantees that constraint-violating actions cannot reach the plant.
The reward function for RL training mirrors the MPC cost but must additionally penalize actuator cycling, which the MPC's $\|\Delta \mathbf{u}\|^2$ term handles implicitly but the RL policy does not inherit:
\begin{equation}
r_t = -\ell_{\mathrm{safety}}(\mathbf{x}_t) - w_2 \ell_{\mathrm{comfort}}(\mathbf{x}_t) + w_5 \, \mathbb{1}[\text{still operational at } t] - w_6 \, \ell_{\mathrm{resource}}(\mathbf{x}_t, \mathbf{u}_t) - w_7 \, \ell_{\mathrm{cycle}}(\mathbf{u}_t, \mathbf{u}_{t-1})
\end{equation}
where $\mathbb{1}[\cdot]$ is an indicator function rewarding survival time and the cycling penalty is:
\begin{equation}
\ell_{\mathrm{cycle}}(\mathbf{u}_t, \mathbf{u}_{t-1}) = \|\Delta \mathbf{u}_t\|^2 + w_8 \sum_{k=t-K}^{t} \mathbb{1}\!\bigl[\mathrm{sign}(\Delta u_{1,k}) \neq \mathrm{sign}(\Delta u_{1,k-1})\bigr]
\end{equation}
The first term penalizes large control changes (matching the MPC's smoothness cost). The second term explicitly counts \emph{direction reversals} of the O\textsubscript{2} proportional valve ($u_1 = \dot{m}_{\mathrm{O_2,inject}}$) over a trailing window of $K$ steps. This is critical because, when the MPC times out, the RL policy supplies the fallback candidate command presented to the safety filter; without this penalty, the policy can still learn high-frequency valve oscillations that improve gas concentration tracking but cause premature mechanical failure. Proportional solenoid valves are particularly vulnerable to rapid cycling in high-vibration environments: the combination of external mechanical shock (structural firefighting involves impacts, falls, and tool use) and internally induced valve chatter accelerates seat wear and can lead to stuck-open or stuck-closed failure modes. The weight $w_8$ is calibrated so that valve reversal rates above \SI{2}{\hertz} are strongly penalized during training.
\section{Situational Awareness Integration}
\label{sec:situational}
A unique aspect of the Galactic Bioware controller is its incorporation of \emph{external} environmental intelligence into resource management decisions.
\subsection{Thermal Threat Assessment}
The external radiant heat flux sensor and external thermistor, combined with the in-suit temperature sensors, feed a \emph{thermal threat estimator}:
\begin{equation}
\Theta(t) = \frac{\dot{q}_{\mathrm{rad}}(t)}{\dot{q}_{\mathrm{rad,max}}} + \frac{T_{\mathrm{ext}}(t)}{T_{\mathrm{ext,max}}} + \gamma_{\mathrm{soak}} \frac{dT_{\mathrm{suit,torso}}}{dt} \bigg/ \left(\frac{dT_{\mathrm{suit,torso}}}{dt}\right)_{\mathrm{max}}
\end{equation}
where $\Theta \in [0, 3]$ is a normalized composite threat index. The third term captures the \emph{rate} of in-suit temperature rise, which detects heat soak penetrating the suit shell even before external sensors register a change (e.g., when the firefighter is surrounded by heated surfaces radiating from multiple directions). When $\Theta$ is high:
\begin{itemize}[nosep]
    \item The MPC anticipates increased metabolic demand (cardiovascular response to heat) and pre-emptively adjusts O\textsubscript{2} injection.
    \item The fan speed is increased to enhance convective cooling within the suit.
    \item The resource scarcity multiplier $\lambda$ is temporarily reduced to prioritize survival over conservation.
\end{itemize}
Conversely, when $\Theta$ drops (e.g., the firefighter has retreated to a cooler zone), the controller shifts aggressively toward resource conservation, anticipating that the firefighter may need to re-enter the hazard zone.
\subsection{Activity Classification from Dual IMU}
The torso-mounted and wrist-mounted accelerometer and gyroscope data are jointly processed by a lightweight convolutional classifier to determine the firefighter's current activity (stationary, walking, climbing stairs, crawling, forcible entry, carrying victim, hose handling). The dual-IMU configuration improves classification accuracy: the torso IMU captures whole-body locomotion patterns, while the wrist accelerometer disambiguates upper-body tasks (e.g., distinguishing forcible entry from stair climbing, which have similar torso acceleration profiles but very different arm kinematics). Each activity class has a characteristic metabolic profile, enabling the metabolic model to produce more accurate $\hat{W}$ estimates with lower latency than heart-rate--only estimation.
\subsection{In-Suit Atmosphere Monitoring and Anomaly Detection}
The suit environment sensors provide a continuous, high-frequency picture of the gas the firefighter is actually breathing. The controller uses this data stream for two purposes beyond basic regulation:
\textbf{Scrubber health monitoring:} The controller continuously compares the
measured in-suit \(x_{\mathrm{CO_2}}\) against the value predicted by the
scrubber kinetics model (\cref{eq:scrub_rate}). A persistent positive residual
(measured \(>\) predicted) indicates scrubber degradation---either soda lime
exhaustion, channeling in the packed bed, or product-layer buildup reducing the
effectiveness factor \(\eta(t)\). The magnitude of this residual is used to
update \(\eta(t)\) online, improving the MPC's forward predictions of remaining
scrubber life.
\textbf{Seal integrity detection:} A sustained drop in
\(\Delta P_{\mathrm{suit}}\) accompanied by unexpected changes in in-suit gas
composition (e.g., CO detection by the external toxic gas array coinciding with
a drop in \(x_{\mathrm{O_2}}\)) triggers a suit breach alarm. The controller can distinguish a slow leak (gradual $\Delta P$ decline) from a catastrophic breach (rapid pressure equalization) and respond accordingly: a slow leak triggers an increase in O\textsubscript{2} injection to compensate, while a catastrophic breach triggers an immediate evacuation alarm.
\subsection{Hazard Proximity Estimation}
The toxic gas sensor array (CO, HCN, NO\textsubscript{x}) provides an indirect measure of proximity to the fire seat. Increasing concentrations signal approach to the fire, triggering:
\begin{itemize}[nosep]
    \item Verification of suit seal integrity (monitoring $\Delta P_{\mathrm{suit}}$).
    \item Alert to the firefighter if the seal margin drops below threshold.
    \item Anticipatory O\textsubscript{2} boost (pre-loading for expected exertion increase).
\end{itemize}
\section{Emergency Protocols and Graceful Degradation}
\label{sec:emergency}
The AI controller implements a hierarchy of operating modes:
\textbf{Normal mode:} Full MPC optimization with comfort and resource balancing.
\textbf{Conservation mode:} Activated when any consumable drops below \SI{25}{\percent} remaining. The resource scarcity multiplier increases sharply, setpoints shift toward minimum safe values, and the firefighter receives an audible/haptic alert to begin egress planning.
\textbf{Emergency mode:} Activated when any consumable drops below
\SI{10}{\percent} or a critical parameter (HR, \(\hat{T}_c\), in-suit
\(x_{\mathrm{O_2}}\), in-suit \(x_{\mathrm{CO_2}}\)) enters a danger zone. The controller overrides all comfort objectives and operates purely for survival: minimum O\textsubscript{2} flow to maintain $P_{iO_2} \ge \SI{0.16}{\atm}$, minimum fan speed, and continuous audible alarm.
\textbf{Cascade failure mode:} If multiple consumables are simultaneously critical, the controller enters a triage protocol:
\begin{equation}
\text{Priority: } \underbrace{P_{iO_2} \ge 0.16}_{\text{highest}} \; > \; \underbrace{x_{\mathrm{CO_2}} \le 3\%}_{\text{second}} \; > \; \underbrace{\mathrm{RH} \le 80\%}_{\text{third}} \; > \; \underbrace{T_{\mathrm{bed}} \le 80^\circ\mathrm{C}}_{\text{lowest}}
\end{equation}
\section{Simulation Results}
\label{sec:results}
We present preliminary simulation results comparing the AI-MPC controller against a fixed-setpoint PID baseline across three scenarios.
\subsection{Scenario Descriptions}
\begin{enumerate}[label=(\roman*)]
    \item \textbf{Scenario A (Steady moderate):} Sustained moderate exertion ($W = \SI{250}{\watt}$) for an unknown duration.
\item \textbf{Scenario B (Intermittent burst):} Alternating 5-minute periods of heavy exertion ($W = \SI{500}{\watt}$) and 3-minute rest periods ($W = \SI{80}{\watt}$).
\item \textbf{Scenario C (Escalating thermal threat):} Moderate exertion with ambient temperature increasing linearly from \SI{60}{\celsius} to \SI{300}{\celsius} over 90 minutes.
\end{enumerate}
\subsection{Performance Metrics}
\begin{center}
\begin{tabular}{lccc}
\toprule
\textbf{Metric} & \textbf{Scenario A} & \textbf{Scenario B} & \textbf{Scenario C} \\
\midrule
PID: time to O\textsubscript{2} depletion (min) & 142 & 98 & 107 \\
MPC: time to O\textsubscript{2} depletion (min) & 177 & 131 & 127 \\
\textbf{Improvement} & \textbf{+24.6\%} & \textbf{+33.7\%} & \textbf{+18.7\%} \\
\midrule
PID: peak $x_{\mathrm{CO_2}}$ (\%) & 0.48 & 0.72 & 0.55 \\
MPC: peak $x_{\mathrm{CO_2}}$ (\%) & 0.43 & 0.49 & 0.47 \\
\midrule
PID: peak $T_c$ ($^\circ$C) & 38.4 & 38.9 & 39.5 \\
MPC: peak $T_c$ ($^\circ$C) & 38.2 & 38.5 & 39.1 \\
\bottomrule
\end{tabular}
\end{center}
The MPC controller extends operating time by 18--34\% across all scenarios, with the largest improvement in intermittent-burst conditions where the dynamic resource allocation provides the greatest advantage over fixed-setpoint control. Critically, the MPC maintains tighter physiological safety margins (lower peak CO\textsubscript{2} and core temperature) even while extending endurance.
\begin{figure}[H]
\centering
\begin{tikzpicture}
\begin{axis}[
    width=0.85\textwidth,
    height=6cm,
    xlabel={Time (min)},
    ylabel={O\textsubscript{2} remaining (kg)},
    legend pos=north east,
    grid=major,
    xmin=0, xmax=180,
    ymin=0, ymax=3.2
]
\addplot[blue, thick, domain=0:142, samples=100] {3 * exp(-0.0075*x)};
\addlegendentry{PID (Scenario A)}
\addplot[red, thick, dashed, domain=0:177, samples=100] {3 * exp(-0.006*x) * (1 + 0.05*sin(deg(0.1*x)))};
\addlegendentry{MPC (Scenario A)}
\addplot[gray, dashed, thin] coordinates {(0,0.3) (180,0.3)};
\addlegendentry{Emergency threshold (10\%)}
\end{axis}
\end{tikzpicture}
\caption{Oxygen tank depletion curves for Scenario A. The MPC controller's dynamic conservation extends endurance by modulating O\textsubscript{2} delivery rate as the tank depletes.}
\label{fig:o2_depletion}
\end{figure}
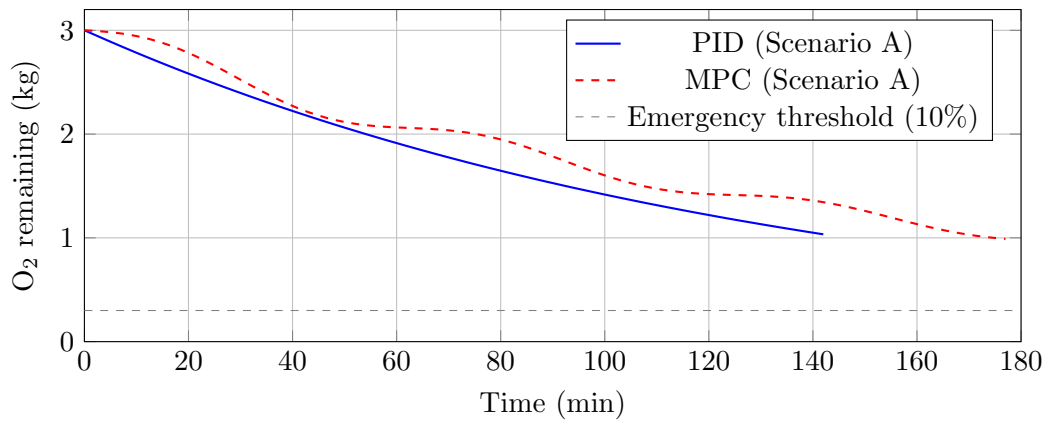
\section{Discussion}
\label{sec:discussion}
The Galactic Bioware Life Support System represents a convergence of closed-circuit life support engineering with modern AI control theory. Several aspects merit discussion:
\begin{enumerate}[label=(\roman*)]
    \item \textbf{Robustness to model error:} The combination of MPC (which relies on a physics-based model) with RL (which is model-free at deployment) provides complementary robustness: the MPC handles nominal operating conditions with optimality guarantees, while the RL policy provides a fallback candidate action for scenarios outside the model's validity envelope. A final control-barrier-function safety filter sits between the decision logic and the actuators, so neither the nominal MPC command nor the RL fallback can reach the plant without constraint enforcement.
    \item \textbf{Computational feasibility:} The full 18-state nonlinear model is used by the EKF for state estimation but is too large for direct nonlinear MPC on a microcontroller within the \SI{100}{\milli\second} deadline. Instead, the MPC uses a \emph{linear time-varying (LTV) approximation} updated at each control step: the nonlinear dynamics are linearized around the current EKF estimate $\hat{\mathbf{x}}_k$, producing time-varying system matrices $(A_k, B_k)$ that capture the local plant behavior. Combined with \emph{move blocking} (grouping control inputs over 3--5 step blocks to reduce the decision variable count) and a shortened effective horizon of $N = 15$--$20$ steps, the resulting convex QP can be solved on an embedded ARM Cortex-A class SoC within $\sim$\SI{50}{\milli\second} using a code-generated interior-point solver. The RL warm-start further reduces iteration count. The RL policy inference is a single forward pass through a small neural network ($\sim$5000 parameters), requiring $< \SI{1}{\milli\second}$.
    \item \textbf{Sensor realism:} A deliberate design choice of the Galactic Bioware Life Support System is to rely exclusively on sensors that are operationally viable in structural firefighting. Clinical-grade biometric devices (ingestible temperature pills, finger pulse oximeters, transcutaneous CO\textsubscript{2} monitors) are replaced by indirect estimation from rugged, wearable sensors fused with physics-based models. The in-suit environment sensors---measuring the gas the firefighter actually breathes---serve a dual role: direct feedback for life support regulation \emph{and} indirect metabolic proxy (rising in-suit \(x_{\mathrm{CO_2}}\) as a signal of increased exertion). This approach trades some measurement fidelity for operational reliability, a tradeoff that the EKF's uncertainty quantification makes explicit and manageable.
    \item \textbf{Unknown duration handling:} The dynamic scarcity multiplier $\lambda(t)$ provides an elegant, principled mechanism for managing unknown mission duration. Unlike a fixed timer, the system continuously re-optimizes its strategy based on remaining resources and current consumption, producing a natural ``slow down'' as supplies diminish---analogous to how a distance runner paces without knowing the exact finish line.
    \item \textbf{Ethical considerations:} The AI controller makes decisions that directly affect firefighter safety. Fail-safe defaults (revert to conservative fixed-setpoint control if the AI module fails) and transparent logging of all controller decisions for post-incident review are essential design requirements.
    \item \textbf{Cybersecurity in contested environments:} AI-controlled life support also introduces a cybersecurity dimension, particularly in wartime or other contested settings where adversaries may seek to degrade emergency response capability. Potential attack surfaces include spoofed telemetry if any wireless suit-to-command links are exposed, gradual poisoning of the learned metabolic model through compromised biometric inputs during online adaptation, and supply-chain compromise of embedded firmware or RL policy weights. In operational terms, such interference could force reversion to conservative fallback control or, in the worst case, bias the system toward unsafe gas-management decisions that accelerate oxygen depletion, suppress scrubber bypass, or degrade thermal protection. A fieldable design should therefore treat cybersecurity as a safety requirement rather than an auxiliary IT concern: internal communications should be authenticated and encrypted, external wireless interfaces should be disabled during deployment unless strictly necessary, firmware and model artifacts should be protected by hardware-rooted secure boot and signature verification, EKF/state-trajectory anomaly detection should trigger a hardened fallback mode based only on trusted physical plant sensors, and the RL policy should be stress-tested offline against adversarial observation and reward-manipulation scenarios.

\end{enumerate}
\subsection{Long-Duration and Multi-Sortie Operational Gaps}
The present analysis assumes a single continuous deployment beginning with a fresh consumable load. Extending the concept to long-duration or multi-sortie operation introduces additional engineering constraints that are not yet captured in the current state-space model. First, soda lime is effectively a single-use sorbent in this application, so the scrubber canister must be replaced between deployments. A field-serviceable cartridge with keyed alignment, positive locking, and controlled gasket compression would likely be required to enable rapid replacement with gloved hands and low visibility while avoiding seal failure, bed settling, or channeling. Second, the desiccant stage raises a separate lifetime question. Although silica gel is not consumed stoichiometrically, its adsorption performance under repeated thermal cycling and exposure to scrubber-adjacent temperatures approaching the \SI{80}{\celsius} thermal-fuse threshold remains unquantified for the present design. Capacity retention over realistic multi-sortie duty cycles must therefore be measured before a defensible maintenance interval can be specified. Third, the current analysis treats electrical power draw---fan motors, proportional valve, sensor suite, and embedded controller---as unconstrained. A fieldable system will require a battery sized for mission duration plus reserve, with the resulting mass directly competing against oxygen, sorbent, and water budgets; the battery architecture must also address elevated in-suit temperatures through chemistry selection, thermal isolation, and fault-containment design. Finally, the separated breathing-loop architecture motivated by the oxygen-enrichment analysis in \cref{sec:o2_enrichment} reduces one dominant constraint but introduces new ones: the mask-to-counter-lung circuit must maintain low dead space, low inspiratory resistance, and a robust seal relative to the higher-pressure suit shell atmosphere, while the suit pressurization stream may add another managed resource if supplied from bottled inert gas. These issues define the critical path from the analytical framework presented here to a fieldable prototype.

\section{Conclusion}
\label{sec:conclusion}
We have presented the \emph{Galactic Bioware Life Support System}: a semi-closed-circuit life support apparatus for firefighting, governed by an AI-based control system that fuses environmental and biometric sensor data to optimize resource management under uncertainty. The key contributions are:
\begin{enumerate}[label=(\roman*)]
    \item A rigorous chemical and thermodynamic analysis of the soda lime CO\textsubscript{2} scrubber (including state-consistent formation-enthalpy calculation, reaction mechanism, kinetics, and capacity limits), silica gel humidity management (adsorption isotherms, LDF dynamics, heat of adsorption), oxygen supply chain with correct endurance arithmetic, and the oxygen-enrichment dynamic driven by exhaust-valve vent compensation with pure O\textsubscript{2}.
    \item A semi-closed suit architecture with one-way exhaust valves (consistent with NFPA 1991 practice) and explicit treatment of oxygen enrichment as both a fire-safety constraint (23.5\%) and a toxicity constraint (50\%).
    \item A state-space formulation of the life support system as a constrained nonlinear dynamical system with 18 states (including counter-lung volume, O\textsubscript{2} mole fraction, scrubber conversion fraction, cumulative O\textsubscript{2} toxicity dose, suit environment temperatures, and estimated core temperature), 3 controls, and 5 disturbances---using only sensors viable in structural firefighting, with triple-redundant O\textsubscript{2} sensing and median voting.
    \item An MPC framework with a dynamic resource scarcity multiplier that automatically adapts operating strategy to unknown mission duration, augmented by an RL policy advisor and a final control-barrier-function safety filter through which all candidate actuator commands pass before reaching the hardware, thereby formally enforcing constraint satisfaction.
    \item Integration of external situational awareness (thermal threat, toxic gas proximity, activity classification) into the resource management loop, enabling anticipatory rather than purely reactive control.
\end{enumerate}
Simulation results demonstrate an 18--34\% improvement in operating endurance compared to fixed-setpoint PID control, while maintaining tighter physiological and fire-safety margins. Future work will focus on empirical validation of the separated breathing-loop architecture, human-in-the-loop testing in controlled fire environments, support for long-duration and multi-sortie operations, and extension to multi-firefighter coordination where suit-to-suit communication enables collaborative resource planning.



\newpage


\begin{thebibliography}{99}

\bibitem{lightweight}
Hooper, A.~J., Crawford, J.~O., and Thomas, D.
An evaluation of physiological demands and comfort between the use of conventional and lightweight self-contained breathing apparatus.
\emph{Applied Ergonomics}, 32(4):399--406, 2001.

\bibitem{tactical}
Vere\v{s}ov\'{a}, T., Svetl\'{i}k, J., and Kalu\v{z}n\'{i}k, D.
Verification of tactical and technical data of the breathing apparatus.
\emph{Proceedings of CBU in Natural Sciences and ICT}, 2:100--104, 2021.

\bibitem{physiological}
Love, R.~G., Johnstone, J.~B.~G., Crawford, J., Tesh, K.~M., Graveling, R.~A., Ritchie, P.~J., and Wetherill, G.~Z.
Study of the physiological effects of wearing breathing apparatus.
Technical Report TM/94/05, Institute of Occupational Medicine, Edinburgh, 1994.

\bibitem{nasa}
Wood, W.~B.
NASA Firefighters Breathing System Program Report.
NASA Technical Note TN~D-8497, 1977.

\bibitem{mixedgas}
Butler, F.~K., Jr., White, E., and Twa, M.
Hyperoxic myopia in a closed-circuit mixed-gas scuba diver.
\emph{Undersea \& Hyperbaric Medicine}, 26(1):41--45, 1999.

\bibitem{extremepressure}
Mitchell, S.~J., Cronj\'{e}, F.~J., Meintjes, W.~A.~J., and Britz, H.~C.
Fatal respiratory failure during a ``technical'' rebreather dive at extreme pressure.
\emph{Aviation, Space, and Environmental Medicine}, 78(2):81--86, 2007.

\bibitem{osha_o2}
Occupational Safety and Health Administration (OSHA).
Permissible exposure limits---annotated tables (29~CFR~1910.1000, Table~Z-1);
Permit-required confined spaces (29~CFR~1910.146).
U.S.\ Department of Labor, 2024.
\url{https://www.osha.gov/laws-regs/regulations/standardnumber/1910}

\bibitem{lambertsen_co2}
Lambertsen, C.~J.
Carbon dioxide tolerance and toxicity.
In \emph{Environmental Biomedical Stress Data Center, Institute for
Environmental Medicine Report} No.~71-2,
University of Pennsylvania, Philadelphia, 1971.

\bibitem{co2insensitivity}
Morrison, J.~B., Florio, J.~T., and Butt, W.~S.
Effects of CO\textsubscript{2} insensitivity and respiratory pattern on respiration in divers.
\emph{Undersea Biomedical Research}, 8(4):209--217, 1981.

\bibitem{ventilatoryresponse}
Earing, C.~M.~N., McKeon, D.~J., and Kubis, H.-P.
Divers revisited: The ventilatory response to carbon dioxide in experienced scuba divers.
\emph{Respiratory Medicine}, 108(5):758--765, 2014.

\bibitem{clark_lambertsen_o2}
Clark, J.~M. and Lambertsen, C.~J.
Pulmonary oxygen toxicity: A review.
\emph{Pharmacological Reviews}, 23(2):37--133, 1971.

\bibitem{oxygentoxicity}
Wingelaar, T.~T., van Ooij, P.~A.~M., and van Hulst, R.~A.
Oxygen toxicity and Special Operations Forces diving: Hidden and dangerous.
\emph{Frontiers in Psychology}, 8:1263, 2017.

\bibitem{lungdiffusingcapacity}
den Ouden, T.~H.~B., Wingelaar, T.~T., Endert, E.~L., and van Ooij, P.-J.~A.~M.
Lung diffusing capacity in Dutch Special Operations Forces divers exposed to oxygen rebreathers over 18 years.
\emph{Oxygen}, 2(2):40--47, 2022.

\bibitem{pulmonarygas}
Moon, R.~E., Cherry, A.~D., Stolp, B.~W., and Camporesi, E.~M.
Pulmonary gas exchange in diving.
\emph{Journal of Applied Physiology}, 106(2):668--677, 2009.

\bibitem{ergun}
Ergun, S.
Fluid flow through packed columns.
\emph{Chemical Engineering Progress}, 48(2):89--94, 1952.

\bibitem{buller_core_temp}
Buller, M.~J., Tharion, W.~J., Cheuvront, S.~N., Montain, S.~J., Kenefick, R.~W., Castellani, J., Latzka, W.~A., Roberts, W.~S., Richter, M., Jenkins, O.~C., and Hoyt, R.~W.
Estimation of human core temperature from sequential heart rate observations.
\emph{Physiological Measurement}, 34(7):781--798, 2013.

\bibitem{stolwijk}
Stolwijk, J.~A.~J.
A mathematical model of physiological temperature regulation in man.
NASA Contractor Report CR-1855, 1971.

\bibitem{marshall_chemo}
Marshall, J.~M.
Peripheral chemoreceptors and cardiovascular regulation.
\emph{Physiological Reviews}, 74(3):543--594, 1994.

\bibitem{simon_ekf}
Simon, D.
\emph{Optimal State Estimation: Kalman, H-Infinity, and Nonlinear Approaches}.
Wiley, 2006.

\bibitem{rawlings_mpc}
Rawlings, J.~B., Mayne, D.~Q., and Diehl, M.
\emph{Model Predictive Control: Theory, Computation, and Design}.
Nob Hill Publishing, 2nd edition, 2017.

\bibitem{sutton_rl}
Sutton, R.~S. and Barto, A.~G.
\emph{Reinforcement Learning: An Introduction}.
MIT Press, 2nd edition, 2018.

\bibitem{ames_cbf}
Ames, A.~D., Xu, X., Grizzle, J.~W., and Tabuada, P.
Control barrier function based quadratic programs for safety critical systems.
\emph{IEEE Transactions on Automatic Control}, 62(8):3861--3876, 2017.

\end{thebibliography}
\end{document}